\documentclass[a4paper,11pt]{article}
\usepackage{jheppub}

\usepackage[T1]{fontenc} 
\usepackage[utf8]{inputenc}
\usepackage{hyperref}
\usepackage{placeins}
\usepackage{amssymb}
\usepackage[english]{babel}
\usepackage{amsmath,amssymb,amsfonts, bm,bbm,slashed, subdepth}
\usepackage{graphicx}
\usepackage[sort&compress]{natbib}
\usepackage{xcolor}
\usepackage[normalem]{ulem}
\usepackage{hyperref}
\usepackage{cleveref}
\definecolor{red}{rgb}{1.0, 0, 0}
\usepackage{enumerate}
\usepackage{epsfig, subfigure}
\usepackage{setspace}
\usepackage{booktabs, tabularx}
\usepackage{units}
\usepackage{placeins}
\usepackage{multirow}
\usepackage{mathtools}

\usepackage{graphicx}
\usepackage{dcolumn}
\usepackage{bm}

\usepackage{placeins} 

\usepackage[english]{babel}
\usepackage{amsmath,amssymb,amsfonts, bm,bbm,slashed, subdepth}
\usepackage{graphicx}
\usepackage[sort&compress]{natbib}
\usepackage{xcolor}
\usepackage[normalem]{ulem}
\usepackage{hyperref}
\usepackage{cleveref}
\definecolor{red}{rgb}{1.0, 0, 0}
\usepackage{enumerate}
\usepackage{epsfig, subfigure}
\usepackage{setspace}
\usepackage{booktabs, tabularx}
\usepackage{units}
\usepackage{placeins}
\usepackage{multirow}
\usepackage{mathtools}

\usepackage{graphicx}
\usepackage{dcolumn}
\usepackage{bm}

\usepackage{placeins} 

\usepackage[colorinlistoftodos,prependcaption,textsize=tiny,obeyFinal]{todonotes}

\newcommand{\drawsquare}[2]{\hbox{%
\rule{#2pt}{#1pt}\hskip-#2pt
\rule{#1pt}{#2pt}\hskip-#1pt
\rule[#1pt]{#1pt}{#2pt}}\rule[#1pt]{#2pt}{#2pt}\hskip-#2pt
\rule{#2pt}{#1pt}}

\def\yzero{\smash{\hbox{$y\kern-4pt\raise1pt\hbox{${}^\circ$}$}}}

\def\ov{\overline}
\def\s2{\frac{1}{\sqrt2}}

\def\beq{\begin{equation}}
	\def\eeq{\end{equation}}
\def\beqa{\begin{eqnarray}}
	\def\eeqa{\end{eqnarray}}

\DeclareMathOperator{\grSU}{SU}
\DeclareMathOperator{\grU}{U}
\DeclareMathOperator{\grUSp}{USp}

\newcommand{\Zbb}{\mathbb{Z}}
\newcommand{\Tbb}{\mathbb{T}}

\newcommand{\fund}{\raisebox{-.5pt}{\drawsquare{6.5}{0.4}}}
\newcommand{\Ysymm}{\raisebox{-.5pt}{\drawsquare{6.5}{0.4}}\hskip-0.4pt%
        \raisebox{-.5pt}{\drawsquare{6.5}{0.4}}}
\newcommand{\Yasymm}{\raisebox{-3.5pt}{\drawsquare{6.5}{0.4}}\hskip-6.9pt%
        \raisebox{3pt}{\drawsquare{6.5}{0.4}}}
\newcommand{\antifund}{\overline{\fund}}

 \renewcommand{\thetable}{\arabic{table}}

	\title{The Complete Search for the Supersymmetric Pati-Salam Models from Intersecting D6-Branes}

\author[a,b,1]{Weikun He}
\author[c,d,2]{Tianjun Li}
\author[a,3]{Rui Sun}

\affiliation[a]{Korea Institute for Advanced Study,\\
		85 Hoegiro, Dongdaemun-Gu, Seoul 02455, Korea}
\affiliation[b]{Academy of Mathematics and Systems Science,\\
		Beijing 100190, P. R. China}
\affiliation[c]{CAS Key Laboratory of Theoretical Physics, Institute of Theoretical Physics,\\
		Chinese Academy of Sciences, Beijing 100190, P. R. China}
\affiliation[d]{School of Physical Sciences, University of Chinese Academy of Sciences,\\
		No.19A Yuquan Road, Beijing 100049, P. R. China}

\emailAdd{heweikun@amss.ac.cn}
\emailAdd{tli@itp.ac.cn}
\emailAdd{sunrui@kias.re.kr}

\abstract{We construct a systematic method   to build all the possible three-family ${\cal N}=1$ supersymmetric Pati-Salam models  from Type IIA orientifolds on  $\Tbb^6/(\mathbb{Z}_2\times \mathbb{Z}_2)$ with intersecting D6-branes, in which the $\grSU(4)_C\times \grSU(2)_L \times \grSU(2)_R $ gauge symmetry can be broken down to	the $\grSU(3)_C \times \grSU(2)_L \times \grU(1)_Y$ Standard Model gauge symmetry by the D-brane splitting 	and supersymmetry preserving Higgs mechanism. This is essentially achieved by solving	all the common solutions for the RR tadpole cancellation conditions, ${\cal N}=1$ supersymmetry conditions, and three generation conditions with deterministic algorithm. We find that there are $202752$ possible supersymmetric Pati-Salam models  in total, and show that there are only $33$ independent models with different gauge coupling relations at string scale after  modding out equivalent relations, such as T-dualities, etc. In particular, there is one and only one independent model which has gauge coupling unification. Furthermore, one can construct  other types of intersecting D-brane models utilizing such deterministic algorithm, and therefore we suggest a brand new method for D-brane model building.
}

\begin{document}
\maketitle


\section{Introduction}

Supersymmetric Pati-Salam models has been investigated intensively with the purpose of realizing ${\cal N}=1$ supersymmetric Standard Models (MSSM) and Standard Models (SM).
It provides consistent constructions of four-dimensional supersymmetric ${\cal N}=1$ chiral models with non-Abelian gauge symmetry on Type II orientifolds for the open string sectors. The chiral fermions on the worldvolume of the D-branes are located at orbifold singularities~\cite{ABPSS, berkooz, ShiuTye, lpt, MCJW, Ibanez, MKRR}, and/or at the intersections of D-branes in the internal space~\cite{bdl} with a T-dual description in terms of magnetized D-branes as shown in~\cite{bachas,urangac}.
Explicit models have been constructed from Type IIA  string theory with orientifolds  $\Tbb^6/(\mathbb{Z}_2\times \mathbb{Z}_2)$  such as in ~\cite{Cvetic:2004ui}. These gauge symmetries are constructed from intersecting D6-branes , with the Pati-Salam gauge symmetries $\grSU(4)_C\times \grSU(2)_L \times \grSU(2)_R$ breaking down to $\grSU(3)_C\times \grSU(2)_L \times \grU(1)_{B-L} \times \grU(1)_{I_{3R}} $ via D6-brane splittings. 
It further breaks down to the SM via the Higgs mechanism.
This provides a way to realize the SM without any additional anomaly-free $\grU(1)$'s around the electroweak scale.
Note that there are also hidden sectors containing $\grUSp(n)$ branes parallel to the orientifold planes or to their ${\Zbb_2}$ images.
These models are normally constructed with at least two confining gauge groups in the hidden sector. For these models, the gaugino condensation triggers supersymmetry breaking and (some) moduli stabilization.
In particular, one of these types of models admits a realistic phenomenology found by in~\cite{Chen:2007px, Chen:2007zu}. 
Its variations have been also visited in~\cite{Chen:2007ms}. 
Moreover,  there are a few other potentially interesting constructions with possible massless 
vector-like fields that might lead to SM~\cite{Cvetic:2004ui}.
These vector fields are not arising from a ${\cal N}=2$ subsector, but can break the Pati-Salam gauge symmetry down to the SM or break $\grU(1)_{B-L}\times \grU(1)_{I_{3R}}$ down to $\grU(1)_Y$.  
For such construction, large wrapping numbers are required because the increased absolute values of the intersection numbers between the $\grU(4)_C$ stack of D-branes and the $\grU(2)_R$ stack (or its orientifold image).
Therefore,  more powerful search methods reaching large wrapping numbers are required.
Many non-supersymmetric three-family SM-like models and generalized unified models have been constructed with intersecting D6-brane models on Type IIA orientifolds, for example in~\cite{bgkl, bkl, afiru}.
These models typically suffer from the large Planck scale corrections at the loop level which results in gauge hierarchy problem.
A large number of the supersymmetric Standard-like models and generalized unified models have been constructed in~\cite{CSU1, CSU2, CP, CPS, CLS1, CLS2, MCIP, CLW, blumrecent, Honecker,
LLG3, Cvetic:2004ui, Cvetic:2004nk, Chen:2005aba, Chen:2005mm, Chen:2005mj,
Chen:2007px, Chen:2007ms, Chen:2007zu, Chen:2008rx, Blumenhagen:2005mu}, solving the aforementioned problem. For a pedagogical introduction to phenomenologically interesting string models constructed with intersecting D-Branes, we refer to~\cite{Blumenhagen:2006ci}.

In \cite{Li:2019nvi, Li:2021},  we constructed supersymmetric Pati-Salam models and generalized Pati-Salam models on Type IIA $\Tbb^6/(\Zbb_2\times \Zbb_2)$  orientifolds with D6-branes intersecting at generic angles to obtain the massless open string state spectra.
In particular, in~\cite{Li:2019nvi}, we systematically studied the three-family ${\cal N}=1$ supersymmetric Pati-Salam model building in Type IIA orientifolds on $\Tbb^6/(\Zbb_2\times \Zbb_2)$ with intersecting D6-branes in which the $\grSU(4)_C\times \grSU(2)_L \times \grSU(2)_R$ gauge symmetries arise from $\grU(n)$ branes. 
We also found that the Type II T-duality in the previous study~\cite{Cvetic:2004ui} is not an equivalence relation in Pati-Salam model building as most of the models are not invariant under $\grSU(2)_L$ and $\grSU(2)_R$ exchange.
In what follows, by swapping the $b-$ and $c-$ stacks of D6-branes, gauge couplings can be redefined and refined, making the gauge coupling unification possible.
The search for supersymmetric Pati-Salam models in the previous works mentioned above are mostly based on random scanning methods.
More recently, powerful machine learning methods have also been employed, such as in \cite{Jim:brane} and \cite{Loges:2021hvn} with realistic intersecting D-brane models revisited.
However, the complete landscape of concrete D-brane models such as supersymmetric Pati-Salam models remains to be drawn.

Distinct from the scanning methods we employed in Ref.~\cite{Li:2019nvi}, in~\cite{Li:2021} by explicitly solving the conditions of  generalized version of  Pati-Salam models, we for the first time systematically discuss the ${\cal N}=1$ supersymmetric
$\grSU(12)_C\times \grSU(2)_L\times \grSU(2)_R$ models, $\grSU(4)_C\times \grSU(6)_L\times \grSU(2)_R$ models, and $\grSU(4)_C\times \grSU(2)_L\times \grSU(6)_R$ models from the Type IIA orientifolds on $\Tbb^6/(\mathbb{Z}_2\times \mathbb{Z}_2)$  with intersecting D6-branes. These gauge symmetries can be broken down to the Pati-Salam gauge symmetry $\grSU(4)_C\times \grSU(2)_L \times \grSU(2)_R$ via three $\grSU(12)_C/\grSU(6)_L/\grSU(6)_R$ adjoint representation Higgs fields, and further down to the Standard Model (SM) via the D-brane splitting  and Higgs mechanism. 

In this paper, we develop a systematic method to obtain the complete landscape of supersymmetric Pati-Salam models.
By solving the RR tadpole cancellation conditions, supersymmetry conditions and three generation conditions step by step, we managed to find a total of all $202752$ possible supersymmetric Pati-Salam models, with $33$ independent gauge coupling relations.
We achieve gauge coupling unification in $6144$ models with three generations obtained from the brane intersection of $a$- and $b$- stacks of branes, of $a$- and $b'$- stack of branes, of $a$- and $c$- stack of branes, and of $a$- and $c'$- stack of branes.
This gauge symmetry can be further broken down to the Standard Model  via D-brane splitting and Higgs mechanism. This method may also be utilized for other string model buildings, especially for brane intersecting models.

The paper is organized as follows. We will first review the basic rules for supersymmetric intersecting D6-brane model building on Type IIA $\Tbb^6/(\Zbb_2\times \Zbb_2)$  orientifolds in Section II and III, and introduce our systematic solving method to obtain all the supersymmetric Pati-Salam models in Section IV.
In Section V, we present one representive  supersymmetric Pati-Salam  model for each of the $33$ different gauge coupling relations.
In Section VI, we discuss the phenomenolgy behaviors of supersymmetric Pati-Salm models.
Finally, Section VII is dedicated to a conclusion and outlook. 

\section{$\Tbb^6 /(\Zbb_2 \times \Zbb_2)$ Orientifolds with Intersecting D6-Branes}
\label{sec:orientifold}
In Refs.~\cite{CSU2, CPS}, supersymmetric models are constructed on Type IIA $\Tbb^6 /(\Zbb_2 \times \Zbb_2)$ orientifolds with D6-branes intersecting at generic angles. 
One can identify the six-torus $\Tbb^{6}$ to the product of  three two-tori  $\Tbb^{6} = \Tbb^{2} \times \Tbb^{2} \times \Tbb^{2}$.
The corresponding complex coordinates on the two-tori are denoted by $z_i$, $i=1,\; 2,\; 3$ respectively.
The generators $\theta$ and $\omega$ of the orbifold group $\Zbb_{2} \times \Zbb_{2}$,
which are respectively associated with the twist vectors $(1/2,-1/2,0)$ and
$(0,1/2,-1/2)$, act on the complex coordinates $z_i$ as below
\beqa
& \theta: & (z_1,z_2,z_3) \mapsto (-z_1,-z_2,z_3)~,~ \nonumber \\
& \omega: & (z_1,z_2,z_3) \mapsto (z_1,-z_2,-z_3)~.~\,
\label{orbifold} \eeqa 
We implement the orientifold projection by gauging the $\Omega R$ symmetry, 
where $\Omega$ is the world-sheet parity, and $R$ is the map defined by
\beqa
R: (z_1,z_2,z_3) \mapsto ({\ov z}_1,{\ov z}_2,{\ov
	z}_3)~.~\, 
\eeqa 
It follows that, we have four kinds of orientifold 6-planes (O6-planes) for each of the actions of $\Omega R$, $\Omega R\theta$, $\Omega R\omega$, and $\Omega R\theta\omega$. 
To cancel the RR charges of O6-planes, stacks of $N_a$ D6-branes wrapping on the factorized three-cycles are introduced.
Also, on each two-torus, there are two possible complex structures consistent with orientifold projection :
rectangular or tilted~\cite{bkl, Chen:2007zu, CSU2,CPS}. 
The homology classes of the three cycles wrapped by the D6-brane stacks can be written as $n_a^i[a_i]+m_a^i[b_i]$ and $n_a^i[a'_i]+m_a^i[b_i]$ for rectangular and tilted tori respectively, with $[a_i']=[a_i]+\frac{1}{2}[b_i]$.
Therefore, in both cases, a generic one cycle can be labelled by two wrapping numbers $(n_a^i,l_a^i)$, 
where $l_{a}^{i}\equiv m_{a}^{i}$ on a rectangular two-torus and $l_{a}^{i}\equiv 2\tilde{m}_{a}^{i}=2m_{a}^{i}+n_{a}^{i}$ on a tilted two-torus. 
As a consequence, $l_a^i-n_a^i$ must be even for a tilted two-torus.

In addition, for $a$- stack of $N_a$ D6-branes  along the cycle $(n_a^i,l_a^i)$, we introduce their $\Omega R$ images as $a'$- stack of $N_a$ D6-branes with wrapping numbers $(n_a^i,-l_a^i)$.
Their homology three-cycles are
\beq
[\Pi_a]=\prod_{i=1}^{3}\left(n_{a}^{i}[a_i]+2^{-\beta_i}l_{a}^{i}[b_i]\right)\quad \text{and} \quad
\left[\Pi_{a'}\right]=\prod_{i=1}^{3}
\left(n_{a}^{i}[a_i]-2^{-\beta_i}l_{a}^{i}[b_i]\right)~,~\, \eeq
where $\beta_i=0$ if the $i$-th torus is rectangular and or $\beta_i=1$ otherwise.
The homology three-cycles wrapped by the four O6-planes, can be written as
\beq \Omega R: [\Pi_{\Omega R}]= 2^3
[a_1]\times[a_2]\times[a_3]~,~\, \eeq \beq \Omega R\omega:
[\Pi_{\Omega
	R\omega}]=-2^{3-\beta_2-\beta_3}[a_1]\times[b_2]\times[b_3]~,~\,
\eeq \beq \Omega R\theta\omega: [\Pi_{\Omega
	R\theta\omega}]=-2^{3-\beta_1-\beta_3}[b_1]\times[a_2]\times[b_3]~,~\,
\eeq \beq
\Omega R\theta:  [\Pi_{\Omega
	R}]=-2^{3-\beta_1-\beta_2}[b_1]\times[b_2]\times[a_3]~.~\,
\label{orienticycles} \eeq 
It follows that the intersection numbers between different stacks can be expressed in terms of wrapping numbers as follows
\beq
I_{ab}=[\Pi_a][\Pi_b]=2^{-k}\prod_{i=1}^3(n_a^il_b^i-n_b^il_a^i)~,~\,
\eeq \beq
I_{ab'}=[\Pi_a]\left[\Pi_{b'}\right]=-2^{-k}\prod_{i=1}^3(n_{a}^il_b^i+n_b^il_a^i)~,~\,
\eeq \beq
I_{aa'}=[\Pi_a]\left[\Pi_{a'}\right]=-2^{3-k}\prod_{i=1}^3(n_a^il_a^i)~,~\,
\eeq \beq {I_{aO6}=[\Pi_a][\Pi_{O6}]=2^{3-k}(-l_a^1l_a^2l_a^3
	+l_a^1n_a^2n_a^3+n_a^1l_a^2n_a^3+n_a^1n_a^2l_a^3)}~,~\,
\label{intersections} \eeq 
where $k=\beta_1+\beta_2+\beta_3$ is the total number of tilted two-tori, 
and $[\Pi_{O6}]=[\Pi_{\Omega R}]+[\Pi_{\Omega R\omega}]+[\Pi_{\Omega
	R\theta\omega}]+[\Pi_{\Omega R\theta}]$ is the sum of four O6-plane
homology three-cycles.
The generic massless particle spectrum for intersecting D6-branes at general angles can be expressed in terms of the intersection numbers as listed in Table \ref{spectrum}, which is valid for both rectangular and tilted two-tori. 
\begin{table}[h] 
	\caption{ 
		General massless particle spectrum for intersecting D6-branes at generic angles. The second column lists the representations of the resulting gauge symmetry group $\grU(N_a/2)$ under the $\Zbb_2\times \Zbb_2$ orbifold projection~\cite{CSU2}.  
The chiral supermultiplets include both scalars and fermions in this supersymmetric constructions. And we choose the positive intersection numbers to denote the left-handed chiral supermultiplets as convention.
	}
	\renewcommand{\arraystretch}{1.25}
	\begin{center}
		\begin{tabular}{|c|c|}
			\hline {\bf Sector} & \phantom{more space inside this box}{\bf
				Representation}
			\phantom{more space inside this box} \\
			\hline\hline
			$aa$   & $\grU(N_a/2)$ vector multiplet  \\
			& 3 adjoint chiral multiplets  \\
			\hline
			$ab+ba$   & $I_{ab}$ $(\fund_a,\antifund_b)$ fermions   \\
			\hline
			$ab'+b'a$ & $I_{ab'}$ $(\fund_a,\fund_b)$ fermions \\
			\hline $aa'+a'a$ &$\frac 12 (I_{aa'} - \frac 12 I_{a,O6})\;\;
			\Ysymm\;\;$ fermions \\
			& $\frac 12 (I_{aa'} + \frac 12 I_{a,O6}) \;\;
			\Yasymm\;\;$ fermions \\
			\hline
		\end{tabular}
	\end{center}
	\label{spectrum}
\end{table}

\subsubsection*{The RR Tadpole Cancellation Conditions}

Moreover, ss discussed in~\cite{Uranga,imr,CSU2}, the tadpole cancellation conditions directly result in the $\grSU(N_a)^3$ cubic non-Abelian anomaly cancellation, and the cancellation of $\grU(1)$ mixed gauge and gravitational anomaly or $[\grSU(N_a)]^2 \grU(1)$ gauge anomaly can be achieved by Green-Schwarz mechanism 
which mediated by untwisted RR fields~\cite{Uranga,imr,CSU2}.
The D6-branes and orientifold O6-planes give rise of RR fields, and thus 
are restricted by the Gauss law in a compact space. To be precise,  
the sum of the RR charges of D6-branes and O6-planes must be zero because of  
the conservations of the RR field flux lines.
The conditions for RR tadpole cancellations can be written as
\begin{eqnarray}
	\sum_a N_a [\Pi_a]+\sum_a N_a
	\left[\Pi_{a'}\right]-4[\Pi_{O6}]=0~,~\,
\end{eqnarray}
where the last terms come from the O6-planes with $-4$ RR charges 
in D6-brane charge unit. 

To make the following discussion less cumbersome, we introduce the following variables which are products of wrapping numbers
\beq
\begin{array}{rrrr}
	A_a \equiv -n_a^1n_a^2n_a^3, & B_a \equiv n_a^1l_a^2l_a^3,
	& C_a \equiv l_a^1n_a^2l_a^3, & D_a \equiv l_a^1l_a^2n_a^3, \\
	\tilde{A}_a \equiv -l_a^1l_a^2l_a^3, & \tilde{B}_a \equiv
	l_a^1n_a^2n_a^3, & \tilde{C}_a \equiv n_a^1l_a^2n_a^3, &
	\tilde{D}_a \equiv n_a^1n_a^2l_a^3.\,
\end{array}
\label{variables}\eeq 
In order to cancel the RR tadpoles, we define an arbitrary number 
of D6-branes wrapping cycles along the orientifold planes, 
which is the so-called ``filler branes'', contributing to the RR tadpole cacellation conditions. Moreover, this also trivially satisfy the four-dimensional ${\cal N}=1$ supersymmetry conditions. The tadpole conditions then can be represented by
\begin{eqnarray}
\label{eq:tadpole}
	-2^k N^{(1)}+\sum_a N_a A_a=-2^k N^{(2)}+\sum_a N_a
	B_a= \nonumber\\ -2^k N^{(3)}+\sum_a N_a C_a=-2^k N^{(4)}+\sum_a
	N_a D_a=-16,\,
\end{eqnarray}
where $2 N^{(i)}$ is the number of filler branes wrapping along the $i$-th O6-plane, as given in Table \ref{orientifold1}.
The filler branes representing the $\grUSp$ group carry the same wrapping numbers 
as one of the O6-planes as shown in Table \ref{orientifold1}. 
We denote the $\grUSp$ group as the $A$-, $B$-, $C$- or $D$-type $\grUSp$ group acccording to non-zero $A$, $B$, $C$ or $D$ filler branes, respectively. 
\renewcommand{\arraystretch}{1.4}
\begin{table}[h] 
	\caption{The wrapping numbers for four O6-planes.} \vspace{0.4cm}
	\begin{center}
		\begin{tabular}{|c|c|c|}
			\hline
			Orientifold Action & O6-Plane & $(n^1,l^1)\times (n^2,l^2)\times
			(n^3,l^3)$\\
			\hline
			$\Omega R$& 1 & $(2^{\beta_1},0)\times (2^{\beta_2},0)\times
			(2^{\beta_3},0)$ \\
			\hline
			$\Omega R\omega$& 2& $(2^{\beta_1},0)\times (0,-2^{\beta_2})\times
			(0,2^{\beta_3})$ \\
			\hline
			$\Omega R\theta\omega$& 3 & $(0,-2^{\beta_1})\times
			(2^{\beta_2},0)\times
			(0,2^{\beta_3})$ \\
			\hline
			$\Omega R\theta$& 4 & $(0,-2^{\beta_1})\times (0,2^{\beta_2})\times
			(2^{\beta_3},0)$ \\
			\hline
		\end{tabular}
	\end{center}
	\label{orientifold1}
\end{table}

\subsubsection*{Conditions for Four-Dimensional $N = 1$ Supersymmetric D6-Brane}
\label{ss:D6brane}

In four-dimensional ${\cal N}=1$ supersymmetric models, $1/4$ supercharges from ten-dimensional Type I T-dual need to be preserved. That is, these $1/4$ supercharges remain preserved under the orientation projection of the intersecting D6-branes and the $\Zbb_2\times \Zbb_2$ orbifold projection on the background manifold. 
One can show that for the four-dimensional ${\cal N}=1$ supersymmetry to survive the orientation projection, the rotation angle of any D6-brane with respect to the orientifold plane must be an element of $\grSU(3)$~\cite{bdl}.
Or equivalently, $\theta_1+\theta_2+\theta_3=0 $ mod $2\pi$, with $\theta_i$ is the angle between the $D6$-brane and orientifold-plane in the $i$-th two-torus. 
Because the $\Zbb_2\times \Zbb_2$ orbifold projection will automatically survive in such D6-brane configuration, 
the four-dimensional ${\cal N}=1$ supersymmetry conditions can be expressed as in~\cite{CPS}
\begin{eqnarray}
	x_A\tilde{A}_a+x_B\tilde{B}_a+x_C\tilde{C}_a+x_D\tilde{D}_a=0,
	\nonumber\\\nonumber \\ A_a/x_A+B_a/x_B+C_a/x_C+D_a/x_D<0,
	\label{susyconditions}
\end{eqnarray} 
where $x_A=\lambda,\;
x_B=\lambda 2^{\beta_2+\beta3}/\chi_2\chi_3,\; x_C=\lambda
2^{\beta_1+\beta3}/\chi_1\chi_3,\; x_D=\lambda
2^{\beta_1+\beta2}/\chi_1\chi_2$, 
and $\chi_i=R^2_i/R^1_i$ are the complex structure moduli for the $i$-th two-torus.  
Here, we introduced the positive parameter $\lambda$ to make all the variables $A,\,B,\,C,\,D$ at equivalent position. 
Because of these conditions, we can classify all possible D6-brane configurations into three types (which preserve the four-dimensional ${\cal N}=1$ supersymmetry): 

(1) If the filler brane has the same wrapping numbers as one of the
O6-planes shown in Table \ref{orientifold1}, the gauge symmetry is $\grUSp$ group. 
In this case, one and only one wrapping number products $A$, $B$, $C$ and $D$ has non-zero and negative value.
According to which one is non-zero, we call the corresponding $\grUSp$ group as the $A$-, $B$-, $C$- or $D$-type $\grUSp$ group, as already mentioned in the last section.

(2) If there is a zero wrapping number, we call the D6-brane a  Z-type D6-brane. In this case, there are two negative and two zero values among $A$, $B$, $C$ and $D$.

(3) If there is a no zero wrapping number, we call the D6-brane a NZ-type D6-brane. Among $A$, $B$, $C$ and $D$,  three of them are negative and one of them is positive.
According to which one is positive, we can classify the NZ-type branes into
the $A$-, $B$-, $C$- and $D$-type NZ branes.
Each type can be further classified into two subtypes with the wrapping numbers taking the form as follows
\begin{eqnarray}
	A1: (-,-)\times(+,+)\times(+,+),~& A2:(-,+)\times(-,+)\times(-,+);\\
	B1: (+,-)\times(+,+)\times(+,+),~& B2:(+,+)\times(-,+)\times(-,+);\\
	C1: (+,+)\times(+,-)\times(+,+),~& C2:(-,+)\times(+,+)\times(-,+);\\
	D1: (+,+)\times(+,+)\times(+,-),~& D2:(-,+)\times(-,+)\times(+,+).
\end{eqnarray}
For the sake of convenience, we will refer to the Z-type and NZ-type D6-branes as $U$-branes since they carry $\grU(n)$ gauge symmetry.

\section{Supersymmetric Pati-Salam Model Building }
\label{sec:models}

To construct SM or SM-like models by intersecting D6-branes model,
in addition to the $\grU(3)_C$ and $\grU(2)_L$ gauge symmetries from stacks of branes,  
we also need at least two extra $\grU(1)$ gauge groups in both supersymmetric and non-supersymmetric models so as to obtain the correct quantum number for right-handed charged leptons~\cite{imr,CSU2,CPS,CP}. 
One is the lepton number symmetry $\grU(1)_L$, while the other is similar to the third component of right-handed weak isospin $\grU(1)_{I_{3R}}$.
Then the hypercharge can be expressed as
\begin{eqnarray}
	Q_Y=Q_{I_{3R}}+{{Q_B-Q_{L}}\over{2}}~,~\,
\end{eqnarray}
where $\grU(1)_B$ is the overall $\grU(1)$ of $\grU(3)_C$.  
In general, the $\grU(1)$ gauge symmetry, coming from a non-Abelian $\grSU(N)$ gauge symmetry, is anomaly free and hence its gauge field is massless.
In our model, $\grU(1)_{B-L}$ and $\grU(1)_{I_{3R}}$ arise respectively from $\grSU(4)_C$ and $\grSU(2)_R$ gauge symmetries. 
Thus, they are both anomaly free and their gauge fields are massless.

We introduce three stacks of D6-branes, called $a$-, $b$-, $c$- stacks with respective D6-brane numbers 8, 4, and 4.
Their respective gauge symmetryies are $\grU(4)_C$, $\grU(2)_L$ and $\grU(2)_R$.
Thus, the Pati-Salam gauge symmetries we obtain is $\grSU(4)_C\times \grSU(2)_L\times \grSU(2)_R$. Via splitting of the D6-branes and Higgs Mechanism, as discussed in~\cite{Cvetic:2004ui}, the Pati-Salam gauge symmetry can be broken down to the SM following the chain
\begin{eqnarray}
	\grSU(4)\times \grSU(2)_L \times \grSU(2)_R  &&
	\xrightarrow[\;a\rightarrow a_1+a_2\;]{}\;  \grSU(3)_C\times \grSU(2)_L
	\times \grSU(2)_R \times \grU(1)_{B-L} \nonumber\\&&
	\xrightarrow[\; c\rightarrow c_1+c_2 \;]{} \; \grSU(3)_C\times \grSU(2)_L\times
	\grU(1)_{I_{3R}}\times \grU(1)_{B-L} \nonumber\\&&
	\xrightarrow[\;\text{Higgs Mechanism}\;]{} \; \grSU(3)_C\times \grSU(2)_L\times \grU(1)_Y~.~\,
\end{eqnarray}
Moreover, to have three families of the SM fermions, the intersection numbers must  satisfy
\begin{eqnarray}
	\label{E3RF} I_{ab} + I_{ab'}~=~3~,~\,\\ \nonumber
 I_{ac} ~=~-3~,~ I_{ac'} ~=~0~.
\end{eqnarray}
The quantum numbers under the $\grSU(4)_C\times \grSU(2)_L\times \grSU(2)_R$ gauge symmetries
are $({\bf 4, 2, 1})$ and $({\bf {\bar 4}, 1, 2})$. 
To satisfy the $I_{ac'} =0 $ condition in \eqref{E3RF}, the stack $a$ D6-branes must be parallel to the orientifold ($\Omega R$) image $c'$ of the $c$-stack along at least one tow-torus.
Instead of \eqref{E3RF}, we could also require the intersection number to satisfy 
\begin{equation}
\label{E3LF} I_{ac}=0~,~I_{ac'}=-3.
\end{equation}
But this actually entirely equivalent due to the symmetry transformation $c\leftrightarrow c'$, the model with intersection numbers  are equivalent to that with $I_{ac}=-3$ and $I_{ac'}=0$. 
The gauge kinetic function for a generic stack $x$ of D6-branes can be expressed as~\cite{CLW}
\begin{eqnarray}
	f_x =  {\bf \textstyle{1\over 4}} \left[ n^1_x n^2_x n^3_x S -
	(\sum_{i=1}^3 2^{-\beta_j-\beta_k}n^i_x l^j_x l^k_x U^i)  \right]
	,\,
\end{eqnarray}
where the real parts of dilaton $S$ and moduli $U^i$ respectively are
\begin{eqnarray}
	{\rm Re}(S) = \frac{M_s^3 R_1^{1} R_1^{2} R_1^{3} }{2\pi g_{s}}~,~\, \\
	{\rm Re}(U^{i}) = {\rm Re}(S)~ \chi_j \chi_k~,~\,
\end{eqnarray}
in which $i\neq j\neq k$, $g_s$ is the string coupling. 
The gauge coupling constant associated with  $x$ can be related with the kinetic function by
\begin{eqnarray}
	\label{g2}
	g_{D6_x}^{-2} &=& |\mathrm{Re}\,(f_x)|.
\end{eqnarray}
By our convention, the holomorphic gauge kinetic functions for $\grSU(4)_C$,  $\grSU(2)_L$
and $\grSU(2)_R$ are respectively identified with the $a$-,  $b$-, and $c$- stacks.
Moreovere, the holomorphic gauge kinetic function for $\grU(1)_Y$ is a linear combination of
those for $\grSU(4)$ and $\grSU(2)_R$.
Namely, by~\cite{bkl, Chen:2007zu}, we have
\begin{equation} \label{fy}
	f_Y =  \frac{3}{5} \,( \frac{2}{3}\, f_{a} + f_{c} ).
\end{equation}
In addition, we can write the tree-level MSSM gauge couplings as
\begin{equation}\label{gy}
	g^2_{a} = \alpha\, g^2_{b} = \beta\, \frac{5}{3}g^2_Y = \gamma\, \left[\pi e^{\phi_4}\right]
\end{equation}
where $g_a^2, g^2_{b}$, and $\frac{5}{3}g^2_Y$ are the strong, weak and hypercharge gauge couplings and $\alpha, \beta, \gamma$ are the ratios between them. 
Moreover, the K\"ahler potential takes the form of
\begin{equation}
	K=-{\rm ln}(S+ \bar S) - \sum_{I=1}^3 {\rm ln}(U^I +{\bar U}^I).~\,
\end{equation}
According to the four-dimensional ${\cal N}=1$ supersymmetry conditions, three stacks of D6-branes, carrying the $\grU(4)_C\times \grU(2)_L \times \grU(2)_R$ gauge symmetry,
generically determine the complex structure moduli $\chi_1$, $\chi_2$ and $\chi_3$.
For this reason,  we only have one independent modulus field.
In order to stabilize these moduli, 
one usually has at least two $\grUSp$ groups with negative $\beta$ functions allowing for gaugino condensations~\cite{Taylor,RBPJS,BDCCM}. 
In general, the one-loop beta function for the $2N^{(i)}$ filler branes, which are on top of $i$-th O6-plane 
and carry $\grUSp(N^{(i)})$ group, is given by~\cite{Cvetic:2004ui}
\begin{eqnarray}
	\beta_i^g&=&-3({N^{(i)}\over2}+1)+2 |I_{ai}|+
	|I_{bi}| +  |I_{ci}|
	+3({N^{(i)}\over2}-1)\nonumber\\
	&=&-6+2 |I_{ai}|+  |I_{bi}|+  |I_{ci}|~,~\,
	\label{betafun}
\end{eqnarray}
which will be considered in our model building also. 

\section{Deterministic Algorithm}
\label{sec:method}

For the searching method we developed in this section, we consider all the above mentioned three generation conditions. Moreover, we do not  restrict the tilted two-torus to be the third one, but consider this can be the first and second two-torus also. Furthermore, we will also include all the possible constructions in terms of wrapping numbers, including the models that can be related via T-duality and D6-brane D6-brane Sign Equivalent Principle as this might lead to models with different gauge couplings as discussed in~\cite{Cvetic:2004ui, Li:2019nvi}.

Recall that each Pati-Salam model is determined by 18 integer parameters, called wrapping numbers $n_a^1,\dotsc, n_c^3, l_a^1,\dotsc, l_c^3$.
These wrapping numbers are required to satisfy the tadpole condition~\eqref{eq:tadpole}, the supersymmetry condition~\eqref{susyconditions} and the three generation conditions~\eqref{E3RF} and \eqref{E3LF}.
Moreover, $l_a^1 - n_a^1$ must be an even number if the corresponding torus a tilted one.
The goal here is to describe a method to find all 18-uplets satisfying these conditions.

Observe that all the conditions in question can be expressed in as polynomial or rational equations or inequalities with the wrapping numbers as variables.
Hence mathematically, the system  consisting of \eqref{eq:tadpole}, \eqref{susyconditions},\eqref{E3RF} and the parity condition  is equivalent to a Diophantine equation.
We shall call this equation the ``Pati-Salam equation''.
On the one hand, there does not exist any algorithm to solve a general Diophantine equation, as implied by the negative solution of Hilbert's tenth problem.
On the other hand, because of the large number of variables, it is unrealistic to use a brute-force search: not to mention that there is a priori no upper bound for the wrapping numbers.
Previously, in~\cite{Li:2019nvi}, random search method has been implemented and some particular solutions to the Pati-Salam equation have been found.

In the present work, we devised a deterministic algorithm which allowed us to find the whole solution set of the Pati-Salam equation.
The solution set happens to be finite, even though we do not have direct and short proof to show why it is so.
Let us describe the main strategy of our algorithm.

First step: we list all the possible combinations of the signs of the twelve wrapping number products $A_a,B_a,\dotsc,C_c,C_d$.
As observed previously in \cite{CPS} and recalled in paragraph~\ref{ss:D6brane}, because of the SUSY condition, $(A_a,B_a,C_a,D_a)$ contains three negative numbers and one positive number, or two negative numbers and two zeroes, or with  one negative number and three zeros. Same applies to the $b-$ stack and the $c-$ stack .

Second step: for each possibility listed in the first step, we append the twelve corresponding inequalities (for example, $A_a>0,B_a<0,C_a<0,D_a<0$) to the our system and try to solve the new system.
Inside the system, we look for the following four kinds of equations or subsystems.
\begin{enumerate}
	\item Equation of the form $v_1 \dotsm v_j =0$ where $v_1,\dotsc,v_j$ are variables. This includes for example $0 = 2I_{ac} = \prod_{i=1}^3(n_a^i l_c^i - l_a^i n_c^i)$ where we view $n_a^i l_c^i - l_a^i n_c^i$, $i=1,2,3$ as variables.
	An equation of this form, obviously, can be solved as $v_1 = 0$ or $v_2=0$ or \dots or $v_j=0$.
	\item Equation of the form $v_1 \dotsm v_j = p$ where $v_1,\dotsc,v_j$ are variables and $p$ is an nonzero integer. Since $p$ has only finitely many factors, an equation of this form can also be solved, leading to finitely many choices for the variables $(v_1,\dotsc,v_j)$.
	\item A system of linear equations of full rank. This includes for example
	\[\left\{\begin{array}{c}
		n_a^1 l_c^1- l_a^1 n_c^1=0\\
		n_a^1 l_c^1 + l_a^1 n_c^1=6,
	\end{array}\right.\]
	where $n_a^1 l_c^1$ and $l_a^1 n_c^1$ are viewed as variables. A system of this kind has unique solution.
	\item A system of linear inequalities which has finitely many integer solutions. Thanks to the first step, it is easier find such system. For example, we could often see subsystem of the following form
	\[\left\{\begin{array}{r}
		4+2A_a + A_b+A_c \geq 0\\
		A_a<0\\
		A_b<0\\
		A_c=0.
	\end{array}\right.\]
	To find such system, we list all linear inequalities in our system. Then for each subset of these inequalities, we can determine whether its integer solution set is finite. Because, a system of linear inequalities over the real numbers correspond to a polyhedron and there there are well-known algorithms (e.g. Fourier-Motzkin elimination) that determine whether the polyhedron has finite volume. If the polyhedron has finite volume, then the system has finitely many integer solutions.
\end{enumerate}
When we find inside our system equations or subsystems of these four kinds, we can solve them with respect to the variables appearing in them. After the resolution, the total number of unknown variables decreases. This might lead to branching into subcases because the solution might not be unique. But there are always only finitely many subcases because we specifically looked for subsystems admitting finite integer solution sets.

Third step: We iterate the Second step for each subcase and repeat.

Fourth step: We stop the iteration when there are no more subsystem of the four kinds.
Now we make use of inequalities of degree larger than 1, such as those given by the SUSY condition~\eqref{susyconditions}. We list all the possible signs of the remaining variables then we determine whether the inequality in question can be achieved. In this way some cases can be eliminated.
For example, say, we are looking for solutions with $A_a<0, B_a<0,C_a<0,D_a>0,A_b<0,B_b<0,C_b=0,D_b=0,A_c<0,B_c>0,C_c<0,D_c<0$ and in the second step we find $n_a^1=1, n_a^2 = -1,n_a^3=1,n_b^1=-1, n_b^2 = -1,n_b^3=1,n_c^1=-1, n_c^2 = -1,n_c^3=1,l_a^1=-1, l_a^3=-1,l_b^1=0,l_c^1=1, l_c^3=2$ as a potential partial solution. The remaining variables are $l_a^2,l_b^2,l_b^3$ and $l_c^2$. Then $B_a<0, B_b<0,B_c>0$ implies that $l_a^2 <0$, $l_c^2 <0$ and $l_b^2 l_b^3>0$. It follows that
\[\frac{x_A}{x_C}= \frac{-3 l_b^2 + l_a^2 l_b^3 + l_b^3 l_c^2}{- l_b^3(l_a^2 + 2 l_c^2)}\]
must be negative, which contradicts the SUSY condition. Thus, this partial solution can be eliminated.

It turns out that these three steps is enough to find all solutions of the Pati-Salam equation. With the help of a computer program, we found in total there are $202752$ different supersymmetric Pati-Salam models with $33$ type of gauge couplings constituting the full landscape. The largest allowed wrapping number is $5$, e.g. as shown in Table \ref{tb:model2} and \ref{tb:model3}. 
In particular, there are $6144$ models with gauge coupling unification at string scale i.e. $g^2_a=g^2_b=g^2_c=(\frac{5}{3}g^2_Y)$. 

\section{Landscape of Supersymmetric Pati-Salam Models}

Among the total possible $202752$ supersymmetric Pati-Salam models, there are $33$ types of gauge coupling relations offered from supersymmetric Pati-Salam models  as shown in Table \ref{tb:model1} - \ref{tb:model33} with one representative model for each type in Appendix~\ref{appA}. For each type of $33$ gauge coupling relations, there are $6144$ models with the tilted two-tori (for which $l_a^i-n_a^i$ must be even) to be the first, second, and third one, while the other two-tori are rectangular.  This can be naturally understood since in our type Type IIA $\Tbb^6 /(\Zbb_2 \times \Zbb_2)$ orientifolds construction, $\Tbb^{6}$ is considered as a six-torus factorized as three two-tori  $\Tbb^{6} = \Tbb^{2} \times \Tbb^{2} \times \Tbb^{2}$, while the first, second, and third two-tori are with equal positions. 
By fixing the titled torus to be the third torus for example, there are $6144/3=2048$ models for each gauge coupling relation.
Moreover, consider the equivalence relations between three generation conditions \eqref{E3RF} and \eqref{E3LF}, there are four types of equivalent three generation relations due to symmetry transformation $b\leftrightarrow b'$ and $c\leftrightarrow c'$. While choosing one type of three generation relation, there will be $2048/4=512$ models for each gauge coupling relation. 

Recall the D6-brane Sign Equivalent Principle(DSEP) discussed in \cite{Cvetic:2004ui}, two D6-brane configurations are equivalent if their wrapping numbers on two arbitrary two-tori have the same magnitude but opposite sign, while their wrapping numbers on the third two-torus are the same. Each stack of D6-brane provides four choices of such construction, and thus DSEP in total contributes $4^3$  types of equivalent models for each gauge coupling relation with $(A_x, \tilde{A}_x ), (B_x, \tilde{B}_x ), (C_x, \tilde{C}_x ), (D_x, \tilde{D}_x )$ in \eqref{variables} remain the same.  Modding out the equivalence relations from DSEP, there are $512/4^3=8$ models for each gauge coupling relations.  

Furthermore, these models with same type of  gauge coupling relation are related via T-dualities as discussed in \cite{Cvetic:2004ui}. In particular, Type I T-dualities perform with $(n_x^j, l_x^j) \rightarrow (-l_x^j, n_x^j),  (n_x^k, l_x^k)\rightarrow (l_x^k, - n_x^k)$ where $x$ runs over all stacks of D6-branes in the model. This leads to 
$(A,\tilde{A})\leftrightarrow(B,\tilde{B}),(C,\tilde{C})\leftrightarrow(D,\tilde{D})$, and further result in the corresponding transformations of moduli parameters $x_A'=x_B, ~x_B'=x_A, ~x_C'=x_D, ~x_D'=x_C$. The $j$-th and $k$-th two-tori can be chosen from these three tori and thus leads to $4$ T-dual types of models for each type of gauge coupling models. In addition, taking into consideration of DSEP, the so-called  type I extended duality(while we choose the third torus to be tilted)  with wrapping number $(n_x^j, l_x^j) \rightarrow (l_x^k, - n_x^k), 
(n_x^k, l_x^k) \rightarrow (-l_x^j, n_x^j)$ provide additional $2$ types of dual models with $j=1, k=2$. 
To sum up, the $8$ models for each type of gauge coupling relation(left by DSEP equivalent relations, etc.) are fully dualized by type I T-duality and type I extended T-duality.  
Thus, there are in total $3\times 4 \times 4^3 \times 4 \times 2= 6144$ equivalent relations. All the possible $202752$ supersymmetric Pati-Salam Models, constituting the supersymmetric  Pati-Salam landscape, provide $33$ physical independent models with $33$ different gauge coupling relations after modding out the equivalent relations.

As we observed in~\cite{Li:2019nvi}, for a three-family supersymmetric Pati-Salam model, there exist  corresponding models by exchanging $b$-stacks and $c$-stacks of D6-branes (with non-trivial T-dualities involved). This corresponds to exchanging the gauge couplings of $\grSU(2)_L$ and $\grSU(2)_R$  at string scale as Model  \ref{tb:model2} and \ref{tb:model3} presented. Note that this construction is not simply swapping the $b$- and $c$-stack of D6-branes, but non-trivial Type II T-dualities also need to be performed. 
In~\cite{Li:2021-2}, we will present that these two models can obtain gauge coupling unification at string scale with certain evolution method.  

There are several classes of models with one, two, three or four $\grUSp$ groups. In particular, such as in Model \ref{tb:model1}, there are four confining $\grUSp(2)$ gauge groups, similar as in Model XVIII in \cite{Li:2019nvi} and Model  I-Z-10 in \cite{Cvetic:2004ui}. Distinct from the random models one can obtain from random scanning, we obtain all the allowed $6144$ supersymmetric Pati-Salam models with gauge coupling unification, i.e. $g^2_a=g^2_b=g^2_c=(\frac{5}{3}g^2_Y)$, where $g^2_a,  g^2_b$, and $\frac{5}{3}g^2_Y$ denote the strong, weak and hypercharge gauge couplings respectively.

We observe there are three classes of  gauge coupling unification models with the large wrapping number $3$ in the first, second, and third torus respectively.  In which the large wrapping number appears in the first, second, and third torus, e.g. in Model \ref{model1}, Model \ref{model2} and Model \ref{model3}. This is easy to understand due to the democratic position of three tori in the orientifold construction $\Tbb^6/\Zbb_2\times \Zbb_2$. For all these $6144$ models with gauge coupling unification, there are four confining $\grUSp(2)$ gauge groups allowing the moduli of these supersymmetric Pati-Salam models stabilized via gaugino condensation.  Moreover, such as Model \ref{model3} and \ref{model4} both with the large wrapping appear in the third torus, are related by swapping their $b-$ and $c-$stacks of branes. As we discussed in \cite{Li:2019nvi} and \cite{Li:2021}, when $g^2_a=g^2_b=g^2_c=(\frac{5}{3}g^2_Y)=4$, the $b-$ and $c-$stacks of branes swapping preserves the gauge coupling unification at string scale. 

Furthermore,  according to the sources of  three generations of the SM fermions, there are four classes of models whose quantum numbers under $\grSU(4)_C\times \grSU(2)_L\times \grSU(2)_R$  with gauge symmetries are $({\bf 4, 2, 1})$ and $({\bf {\overline{4}}, 1, 2})$ in supersymmetric Pati-Salam construction.  For these  models in Appendix~\ref{appB}, e.g. Model~\ref{model1} with $I_{ab}=3$ and $I_{ac'} =-3$, Model~\ref{model4} with $I_{ab'}=3$ and $I_{ac'} =-3$ , Model~\ref{model5} with $I_{ab'}=3$ and $I_{ac} =-3$, Model~\ref{model6} with $I_{ab}=3$ and $I_{ac} =-3$ achieve three generations of the SM fermions.

\section{Phenomenological Studies}
\label{sec:pheno}

In this section, we shall discuss the phenomenological features of our models.  In these four models in Appendix~\ref{appB}, the gauge symmetry is $\grU(4)\times \grU(2)_L\times \grU(2)_R\times \grUSp(2)^4$.  The $\beta$ functions of $\grUSp(2)$ group are negative, with four confining gauge groups  in the hidden sector, thus we can break supersymmety via gaugino condensation. For the other type of models shown in Appendix~\ref{appA}, there are models with one, two, three and four confining groups. 
For these models with two confining $\grUSp(N)$ gauge groups, one can perform a general analysis of  the non-perturbative superpotential with tree-level gauge couplings, there can exist extrema with the stabilizations of dilaton and complex structure moduli as discussed in~\cite{CLW}. 
However, these extrema might be saddle points and thus will not break the supersymmetry. If the models have three or four confining $\grUSp(N)$ gauge groups,  the non-perturbative superpotiential allows for moduli stabilization and supersymmetry breaking at the stable extremum in general. 

Now we take Models \ref{model4} as examples to show explicitly the full spectrum of the models with gauge coupling unification from Appendix~\ref{appB} in Table \ref{spmodel4}. At the near string scale, we have  $\grSU(3)_C \times \grSU(2)_L\times \grU(1)_Y$  gauge coupling unification. Its  composite states are shown explicitly in Table \ref{Composite Particles model4}.
\begin{table}
	[h] \scriptsize
	\renewcommand{\arraystretch}{0.6}
	\caption{Chiral spectrum in the open string sector - Model \ref{model4}} \label{spmodel4}
	\begin{center}
		\begin{tabular}{c|ccccccc}\hline
			& $\grSU(4)\times \grSU(2)_L$
			& & & & & & \\
			Model $2$ 		& $\times \grSU(2)_R\times \grUSp(2)^4$
			& $Q_4$ & $Q_{2L}$ & $Q_{2R}$ & $Q_{em}$ & $B-L$ & Field \\
			\hline\hline
			$ab'$ & $3\times (4, 2,1,1,1,1,1)$ & 1 & 1 & 0  & $-\frac 13, \frac 23,-1, 0$ & $\frac 13,-1$ & $Q_L, L_L$\\
			$ac'$ & $3\times (\overline{4},1,\overline{2},1,1,1,1)$ & -1 & 0 & $-1$& $\frac 13, -\frac 23,1, 0$ & $-\frac 13,1$ & $Q_R, L_R$\\
			$a2$ & $1\times (4,1,1,1,2,1,1)$ & $1$ & 0 & 0 & $\frac 16,-\frac 12$ & $\frac 13,-1$ & \\
			$a3$ & $1\times (4,1,1,1,1,2,1)$ & 1 & 0 & 0   & $\frac 16,-\frac 12$ & $\frac 13,-1$ & \\
		$b1$ & $3\times(1,2,1,2,1,1,1)$ & 0 & 1 & 0   & $\pm \frac 12$ & 0 & \\
		$b3$ & $1\times(1,2,1,1,1,2,1)$ & 0 & 1 & 0   & $\pm  \frac 12$ & 0 & \\
		$c2$ & $1\times(1,1,2,1,2,1,1)$ & 0 & 0 & 1   & $\pm \frac 12$ & 0 & \\
		$c4$ & $3\times(1,1,2,1,1,1,2 )$ & 0 & 0 & -1   & $\pm \frac 12$ & 0 & \\
			$b_{\Ysymm}$ & $2\times(1,\overline{3},1,1,1,1,1)$ & 0 & -2 & 0   & $0,\pm 1$ & 0 & \\
			$b_{\overline{\Yasymm}}$ & $2\times(1,1,1,1,1,1,1)$ & 0 & 2 & 0   & 0 & 0 & \\
			$c_{\overline{\Ysymm}}$ & $2\times(1,1,3,1,1,1,1)$ & 0 & 0 & 2   & $0,\pm 1$ & 0 & \\
			$c_{\Yasymm}$ & $2\times(1,1,1,1,1,1,1)$ & 0 & 0 & -2   & 0 & 0 & \\
			\hline
		\end{tabular}
	\end{center}
\end{table}

\begin{table}
	[h] \scriptsize
	\renewcommand{\arraystretch}{1.0}
	\caption{The composite particle spectrum of Model~\ref{model4}, which is
		formed due to the strong forces from hidden sector.}
	\label{Composite Particles model4}
	\begin{center}
		\begin{tabular}{cc c c}\hline
			\multicolumn{2}{c }{Model~\ref{model4}} &
			\multicolumn{2}{c}{$\grSU(4)\times \grSU(2)_L\times \grSU(2)_R \times
				\grUSp(2)^4$} \\
			\hline Confining Force & Intersection & Exotic Particle
			Spectrum & Confined Particle Spectrum \\
			\hline
			$\grUSp(2)_1$ &$b1$ & $3\times (1,2,1,2,1,1,1)$ & $6\times (1,2^2,1,1,1,1,1)$\\
			\hline
			$\grUSp(2)_2$ &$a2$ & $1\times (4,1,1,1,2,1,1)$ & $1\times (4^2,1,1,1,1,1,1)$, $1\times(4,1,2,1,1,1,1)$\\
			&$c2$ & $1\times(1,1,2,1,2,1,1)$ &  $1\times(1,1,2^2,1,1,1,1)$\\
			\hline
			$\grUSp(2)_3$ &$a3$ & $1\times (4,1,1,1,1,2,1)$ & $1\times (4^2,1,1,1,1,1,1)$, $1\times(4,2,1,1,1,1,1)$\\
			&$b3$ & $1\times(1,2,1,1,1,2,1)$ &  $1\times(1,2^2,1,1,1,1,1)$\\
			\hline
			$\grUSp(2)_4$ &$c4$ & $3\times(1,1,2,1,1,1,2)$ & $6\times(1,1,2^2,1,1,1,1)$\\
			\hline
		\end{tabular}
	\end{center}
\end{table}

Model~\ref{model4} has four confining gauge groups. Thereinto, both $\grUSp(2)_2$
and $\grUSp(2)_3$ have two charged intersections. Thus for them, besides self-confinement, the mixed-confinement between different intersections also exist, which yields the chiral supermultiplets $(4,1,2,1,1,1,1)$ and $(4,1,1,1,1,2,1)$. As for $\grUSp(2)_1$ and $\grUSp(2)_4$,
they have only one charged intersection. Therefore, there is no mixed-confinement, their self-confinement leads to 6 tensor representations for each of them.
In addition, one can check from the spectra that no new anomaly is introduced to the
remaining gauge symmetry, namely, such models are  anomaly-free~\cite{CLS1}.


\section{Conclusions and Outlook}

In this paper, we developed a systematic method for supersymmetric Pati-Salam models, and managed to obtain all the possible models from this construction. In such a way, we complete the landscape of supersymmetric Pati-Salam models constructed from Type IIA  string theory with orientifolds  $\Tbb^6/(\mathbb{Z}_2\times \mathbb{Z}_2)$ for the first time. 

We found that there are in total $33$ types of allowed gauge coupling relations for  supersymmetric Pati-Salam models, with in total $6144$ models with gauge coupling unification among total $202752$ models. These are all the allowed models by solving the common solutions of RR tadpole cancellation conditions, supersymmetry conditions, and three generation conditions simultaneously. Thus all the possible models are found. The largest allowed wrapping number is $5$ in this system. 
Took one model with gauge coupling unification at near string scale as example, we studied their phenomenology features, and observe well anomaly-free behaviors with self- and mixed-confinements.

Distinct from the earlier investigations, we considered all the intersection conditions to have three families of the SM fermions, and allow any of the three tori to be tilted. Incorporated with T-dualities, type II T-dualities and D6-brane Sign Equivalent Principle, we obtained all the possible constructed supersymmetric Pati-Salam models. 

We expect that the searching method for landscape of supersymmetric Pati-Salam models can be generalized to other string model buildings, especially brane intersecting models with RR tadpole cancellation conditions, supersymmetry conditions, and three generation conditions required simultaneously.

\FloatBarrier
\begin{acknowledgments}
	TL is supported by the National Key Research and Development Program of China Grant No. 2020YFC2201504, as well as by the Projects No. 11875062 and No. 11947302 supported by the National Natural Science Foundation of China. WH is supported by KIAS Individual Grant MG080401. RS is supported by KIAS Individual Grant PG080701.   
	
\end{acknowledgments}


\appendix

\FloatBarrier
\section{Independent Supersymmetric Pati-Salam Models}
\label{appA}

In this section, we present the supersymmetric Pati-Salam models with $33$ types of allowed gauge coupling relations on the landscape of supersymmetric Pati-Salam model building{\color{blue}\footnote{The full data of wrapping numbers $(n_x^i, l_x^i)$ with $x=a, b, c$ and $i=1,2,3$ for 202752 models are listed in http://newton.kias.re.kr/~sunrui/files/finaldata.csv.}}. 

\begin{table}[h]\scriptsize
	\caption{D6-brane configurations and intersection numbers of Model \ref{tb:model1}, and its MSSM gauge coupling relation is $g^2_a=g^2_b=g^2_c=(\frac{5}{3}g^2_Y)=4 \sqrt{\frac{2}{3}} \pi  e^{\phi ^4}$.}
	\label{tb:model1}
	\begin{center}
\begin{tabular}{|c||c|c||c|c|c|c|c|c|c|c|c|c|}
			\hline\rm{model} \ref{tb:model1} & \multicolumn{12}{c|}{$U(4)\times U(2)_L\times U(2)_R\times USp(2)^4 $}\\
			\hline \hline			\rm{stack} & $N$ & $(n^1,l^1)\times(n^2,l^2)\times(n^3,l^3)$ & $n_{\Ysymm}$& $n_{\Yasymm}$ & $b$ & $b'$ & $c$ & $c'$ & 1 & 2 & 3 & 4\\
			\hline
			$a$ & 8 & $(1,1)\times (1,0)\times (1,-1)$ & 0 & 0  & 3 & 0 & 0 & -3 & 0 & 1 & 0 & -1\\
			$b$ & 4 & $(-1,0)\times (-1,3)\times (1,1)$ & 2 & -2  & - & - & 0 & 0 & 0 & 0 & -3 & 1\\
			$c$ & 4 & $(0,1)\times (-1,3)\times (-1,1)$ & 2 & -2  & - & - & - & - & -3 & 1 & 0 & 0\\
			\hline
			1 & 2 & $(1, 0)\times (1, 0)\times (2, 0)$& \multicolumn{10}{c|}{$x_A = \frac{1}{3}x_B = x_C = \frac{1}{3}x_D$}\\
			2 & 2 & $(1, 0)\times (0, -1)\times (0, 2)$& \multicolumn{10}{c|}{$\beta^g_1=-3$, $\beta^g_2=-3$, $\beta^g_3=-3$, $\beta^g_4=-3$}\\
			3 & 2 & $(0, -1)\times (1, 0)\times (0, 2)$& \multicolumn{10}{c|}{$\chi_1=1$, $\chi_2=\frac{1}{3}$, $\chi_3=2$}\\
			4 & 2 & $(0, -1)\times (0, 1)\times (2, 0)$& \multicolumn{10}{c|}{}\\
			\hline
		\end{tabular}
	\end{center}
\end{table}
\begin{table}\scriptsize
	\caption{D6-brane configurations and intersection numbers of Model \ref{tb:model2}, and its MSSM gauge coupling relation is $g^2_a=\frac{7}{6}g^2_b=\frac{5}{6}g^2_c=\frac{25}{28}(\frac{5}{3}g^2_Y)=\frac{8}{27} 5^{3/4} \sqrt{7} \pi  e^{\phi ^4}$.}
	\label{tb:model2}
	\begin{center}
		\begin{tabular}{|c||c|c||c|c|c|c|c|c|c|c|}
			\hline\rm{model} \ref{tb:model2} & \multicolumn{10}{c|}{$U(4)\times U(2)_L\times U(2)_R\times USp(2)^2 $}\\
			\hline \hline			\rm{stack} & $N$ & $(n^1,l^1)\times(n^2,l^2)\times(n^3,l^3)$ & $n_{\Ysymm}$& $n_{\Yasymm}$ & $b$ & $b'$ & $c$ & $c'$ & 1 & 4\\
			\hline
			$a$ & 8 & $(1,-1)\times (-1,1)\times (1,-1)$ & 0 & 4  & 0 & 3 & 0 & -3 & -1 & 1\\
			$b$ & 4 & $(0,1)\times (-2,1)\times (-1,1)$ & -1 & 1  & - & - & 0 & -1 & -1 & 0\\
			$c$ & 4 & $(-1,0)\times (5,2)\times (-1,1)$ & 3 & -3  & - & - & - & - & 0 & -5\\
			\hline
			1 & 2 & $(1, 0)\times (1, 0)\times (2, 0)$& \multicolumn{8}{c|}{$x_A = 2x_B = \frac{14}{5}x_C = 7x_D$}\\
			4 & 2 & $(0, -1)\times (0, 1)\times (2, 0)$& \multicolumn{8}{c|}{$\beta^g_1=-3$, $\beta^g_4=1$}\\
			& & & \multicolumn{8}{c|}{$\chi_1=\frac{7}{\sqrt{5}}$, $\chi_2=\sqrt{5}$, $\chi_3=\frac{4}{\sqrt{5}}$}\\
			\hline
		\end{tabular}
	\end{center}
\end{table}
\begin{table}\scriptsize
	\caption{D6-brane configurations and intersection numbers of Model \ref{tb:model3}, and its MSSM gauge coupling relation is $g^2_a=\frac{5}{6}g^2_b=\frac{7}{6}g^2_c=\frac{35}{32}(\frac{5}{3}g^2_Y)=\frac{8}{27} 5^{3/4} \sqrt{7} \pi  e^{\phi ^4}$.}
	\label{tb:model3}
	\begin{center}
		\begin{tabular}{|c||c|c||c|c|c|c|c|c|c|c|}
			\hline\rm{model} \ref{tb:model3} & \multicolumn{10}{c|}{$U(4)\times U(2)_L\times U(2)_R\times USp(2)^2 $}\\
			\hline \hline			\rm{stack} & $N$ & $(n^1,l^1)\times(n^2,l^2)\times(n^3,l^3)$ & $n_{\Ysymm}$& $n_{\Yasymm}$ & $b$ & $b'$ & $c$ & $c'$ & 1 & 4\\
			\hline
			$a$ & 8 & $(-1,-1)\times (1,1)\times (1,1)$ & 0 & -4  & 0 & 3 & 0 & -3 & 1 & -1\\
			$b$ & 4 & $(-5,2)\times (-1,0)\times (1,1)$ & -3 & 3  & - & - & 0 & 1 & 0 & 5\\
			$c$ & 4 & $(-2,-1)\times (0,1)\times (1,1)$ & 1 & -1  & - & - & - & - & 1 & 0\\
			\hline
			1 & 2 & $(1, 0)\times (1, 0)\times (2, 0)$& \multicolumn{8}{c|}{$x_A = \frac{14}{5}x_B = 2x_C = 7x_D$}\\
			4 & 2 & $(0, -1)\times (0, 1)\times (2, 0)$& \multicolumn{8}{c|}{$\beta^g_1=-3$, $\beta^g_4=1$}\\
			& & & \multicolumn{8}{c|}{$\chi_1=\sqrt{5}$, $\chi_2=\frac{7}{\sqrt{5}}$, $\chi_3=\frac{4}{\sqrt{5}}$}\\
			\hline
		\end{tabular}
	\end{center}
\end{table}
\begin{table}\scriptsize
	\caption{D6-brane configurations and intersection numbers of Model \ref{tb:model4}, and its MSSM gauge coupling relation is $g^2_a=\frac{5}{3}g^2_b=g^2_c=(\frac{5}{3}g^2_Y)=4 \sqrt{\frac{2}{3}} \pi  e^{\phi ^4}$.}
	\label{tb:model4}
	\begin{center}
		\begin{tabular}{|c||c|c||c|c|c|c|c|c|c|c|}
			\hline\rm{model} \ref{tb:model4} & \multicolumn{10}{c|}{$U(4)\times U(2)_L\times U(2)_R\times USp(4)^2 $}\\
			\hline \hline			\rm{stack} & $N$ & $(n^1,l^1)\times(n^2,l^2)\times(n^3,l^3)$ & $n_{\Ysymm}$& $n_{\Yasymm}$ & $b$ & $b'$ & $c$ & $c'$ & 2 & 4\\
			\hline
			$a$ & 8 & $(1,1)\times (1,0)\times (1,-1)$ & 0 & 0  & 3 & 0 & 0 & -3 & 1 & -1\\
			$b$ & 4 & $(-2,1)\times (-1,1)\times (1,1)$ & -2 & -6  & - & - & 4 & 0 & -1 & 2\\
			$c$ & 4 & $(0,1)\times (-1,3)\times (-1,1)$ & 2 & -2  & - & - & - & - & 1 & 0\\
			\hline
			2 & 4 & $(1, 0)\times (0, -1)\times (0, 2)$& \multicolumn{8}{c|}{$x_A = \frac{1}{3}x_B = x_C = \frac{1}{3}x_D$}\\
			4 & 4 & $(0, -1)\times (0, 1)\times (2, 0)$& \multicolumn{8}{c|}{$\beta^g_2=-2$, $\beta^g_4=-2$}\\
			& & & \multicolumn{8}{c|}{$\chi_1=1$, $\chi_2=\frac{1}{3}$, $\chi_3=2$}\\
			\hline
		\end{tabular}
	\end{center}
\end{table}
\begin{table}\scriptsize
	\caption{D6-brane configurations and intersection numbers of Model \ref{tb:model5}, and its MSSM gauge coupling relation is $g^2_a=\frac{9}{5}g^2_b=g^2_c=(\frac{5}{3}g^2_Y)=\frac{12}{5} \sqrt{2} \pi  e^{\phi ^4}$.}
	\label{tb:model5}
	\begin{center}
		\begin{tabular}{|c||c|c||c|c|c|c|c|c|c|c|c|c|}
			\hline\rm{model} \ref{tb:model5} & \multicolumn{12}{c|}{$U(4)\times U(2)_L\times U(2)_R\times USp(2)^4 $}\\
			\hline \hline			\rm{stack} & $N$ & $(n^1,l^1)\times(n^2,l^2)\times(n^3,l^3)$ & $n_{\Ysymm}$& $n_{\Yasymm}$ & $b$ & $b'$ & $c$ & $c'$ & 1 & 2 & 3 & 4\\
			\hline
			$a$ & 8 & $(1,1)\times (1,0)\times (1,-1)$ & 0 & 0  & 2 & 1 & 0 & -3 & 0 & 1 & 0 & -1\\
			$b$ & 4 & $(-1,0)\times (-1,1)\times (1,3)$ & -2 & 2  & - & - & 4 & 4 & 0 & 0 & -1 & 3\\
			$c$ & 4 & $(0,1)\times (-1,3)\times (-1,1)$ & 2 & -2  & - & - & - & - & -3 & 1 & 0 & 0\\
			\hline
			1 & 2 & $(1, 0)\times (1, 0)\times (2, 0)$& \multicolumn{10}{c|}{$x_A = \frac{1}{3}x_B = \frac{1}{9}x_C = \frac{1}{3}x_D$}\\
			2 & 2 & $(1, 0)\times (0, -1)\times (0, 2)$& \multicolumn{10}{c|}{$\beta^g_1=-3$, $\beta^g_2=-3$, $\beta^g_3=-5$, $\beta^g_4=-1$}\\
			3 & 2 & $(0, -1)\times (1, 0)\times (0, 2)$& \multicolumn{10}{c|}{$\chi_1=\frac{1}{3}$, $\chi_2=1$, $\chi_3=\frac{2}{3}$}\\
			4 & 2 & $(0, -1)\times (0, 1)\times (2, 0)$& \multicolumn{10}{c|}{}\\
			\hline
		\end{tabular}
	\end{center}
\end{table}
\begin{table}\scriptsize
	\caption{D6-brane configurations and intersection numbers of Model \ref{tb:model6}, and its MSSM gauge coupling relation is $g^2_a=2g^2_b=g^2_c=(\frac{5}{3}g^2_Y)=\frac{16 \sqrt[4]{2} \pi  e^{\phi ^4}}{3 \sqrt{3}}$.}
	\label{tb:model6}
	\begin{center}
		\begin{tabular}{|c||c|c||c|c|c|c|c|c|c|c|c|}
			\hline\rm{model} \ref{tb:model6} & \multicolumn{11}{c|}{$U(4)\times U(2)_L\times U(2)_R\times USp(2)^3 $}\\
			\hline \hline			\rm{stack} & $N$ & $(n^1,l^1)\times(n^2,l^2)\times(n^3,l^3)$ & $n_{\Ysymm}$& $n_{\Yasymm}$ & $b$ & $b'$ & $c$ & $c'$ & 2 & 3 & 4\\
			\hline
			$a$ & 8 & $(1,1)\times (1,0)\times (1,-1)$ & 0 & 0  & 3 & 0 & 0 & -3 & 1 & 0 & -1\\
			$b$ & 4 & $(-1,0)\times (-2,3)\times (1,1)$ & 1 & -1  & - & - & 3 & 0 & 0 & -3 & 2\\
			$c$ & 4 & $(0,1)\times (-1,3)\times (-1,1)$ & 2 & -2  & - & - & - & - & 1 & 0 & 0\\
			\hline
			2 & 2 & $(1, 0)\times (0, -1)\times (0, 2)$& \multicolumn{9}{c|}{$x_A = \frac{1}{3}x_B = \frac{1}{2}x_C = \frac{1}{3}x_D$}\\
			3 & 2 & $(0, -1)\times (1, 0)\times (0, 2)$& \multicolumn{9}{c|}{$\beta^g_2=-3$, $\beta^g_3=-3$, $\beta^g_4=-2$}\\
			4 & 2 & $(0, -1)\times (0, 1)\times (2, 0)$& \multicolumn{9}{c|}{$\chi_1=\frac{1}{\sqrt{2}}$, $\chi_2=\frac{\sqrt{2}}{3}$, $\chi_3=\sqrt{2}$}\\
			\hline
		\end{tabular}
	\end{center}
\end{table}
\begin{table}\scriptsize
	\caption{D6-brane configurations and intersection numbers of Model \ref{tb:model7}, and its MSSM gauge coupling relation is $g^2_a=g^2_b=\frac{5}{3}g^2_c=\frac{25}{19}(\frac{5}{3}g^2_Y)=4 \sqrt{\frac{2}{3}} \pi  e^{\phi ^4}$.}
	\label{tb:model7}
	\begin{center}
		\begin{tabular}{|c||c|c||c|c|c|c|c|c|c|c|}
			\hline\rm{model} \ref{tb:model7} & \multicolumn{10}{c|}{$U(4)\times U(2)_L\times U(2)_R\times USp(4)^2 $}\\
			\hline \hline			\rm{stack} & $N$ & $(n^1,l^1)\times(n^2,l^2)\times(n^3,l^3)$ & $n_{\Ysymm}$& $n_{\Yasymm}$ & $b$ & $b'$ & $c$ & $c'$ & 2 & 4\\
			\hline
			$a$ & 8 & $(1,1)\times (1,0)\times (1,-1)$ & 0 & 0  & 3 & 0 & 0 & -3 & 1 & -1\\
			$b$ & 4 & $(1,0)\times (1,-3)\times (1,1)$ & 2 & -2  & - & - & -4 & 0 & 0 & 1\\
			$c$ & 4 & $(1,2)\times (-1,1)\times (-1,1)$ & -2 & -6  & - & - & - & - & 2 & -1\\
			\hline
			2 & 4 & $(1, 0)\times (0, -1)\times (0, 2)$& \multicolumn{8}{c|}{$x_A = \frac{1}{3}x_B = x_C = \frac{1}{3}x_D$}\\
			4 & 4 & $(0, -1)\times (0, 1)\times (2, 0)$& \multicolumn{8}{c|}{$\beta^g_2=-2$, $\beta^g_4=-2$}\\
			& & & \multicolumn{8}{c|}{$\chi_1=1$, $\chi_2=\frac{1}{3}$, $\chi_3=2$}\\
			\hline
		\end{tabular}
	\end{center}
\end{table}
\begin{table}\scriptsize
	\caption{D6-brane configurations and intersection numbers of Model \ref{tb:model8}, and its MSSM gauge coupling relation is $g^2_a=\frac{11}{6}g^2_b=\frac{5}{6}g^2_c=\frac{25}{28}(\frac{5}{3}g^2_Y)=\frac{8 \sqrt[4]{2} 5^{3/4} \pi  e^{\phi ^4}}{7 \sqrt{3}}$.}
	\label{tb:model8}
	\begin{center}
		\begin{tabular}{|c||c|c||c|c|c|c|c|c|c|}
			\hline\rm{model} \ref{tb:model8} & \multicolumn{9}{c|}{$U(4)\times U(2)_L\times U(2)_R\times USp(2) $}\\
			\hline \hline			\rm{stack} & $N$ & $(n^1,l^1)\times(n^2,l^2)\times(n^3,l^3)$ & $n_{\Ysymm}$& $n_{\Yasymm}$ & $b$ & $b'$ & $c$ & $c'$ & 3\\
			\hline
			$a$ & 8 & $(1,1)\times (1,0)\times (1,-1)$ & 0 & 0  & 3 & 0 & 0 & -3 & 0\\
			$b$ & 4 & $(2,-1)\times (1,-1)\times (1,1)$ & -2 & -6  & - & - & 8 & 0 & -2\\
			$c$ & 4 & $(-2,5)\times (0,1)\times (-1,1)$ & 3 & -3  & - & - & - & - & 2\\
			\hline
			3 & 2 & $(0, -1)\times (1, 0)\times (0, 2)$& \multicolumn{7}{c|}{$x_A = \frac{1}{6}x_B = \frac{2}{5}x_C = \frac{1}{6}x_D$}\\
			& & & \multicolumn{7}{c|}{$\beta^g_3=-2$}\\
			& & & \multicolumn{7}{c|}{$\chi_1=\sqrt{\frac{2}{5}}$, $\chi_2=\frac{\sqrt{\frac{5}{2}}}{6}$, $\chi_3=2 \sqrt{\frac{2}{5}}$}\\
			\hline
		\end{tabular}
	\end{center}
\end{table}
\begin{table}\scriptsize
	\caption{D6-brane configurations and intersection numbers of Model \ref{tb:model9}, and its MSSM gauge coupling relation is $g^2_a=\frac{10}{9}g^2_b=\frac{4}{9}g^2_c=\frac{4}{7}(\frac{5}{3}g^2_Y)=\frac{16}{27} 2^{3/4} \sqrt{5} \pi  e^{\phi ^4}$.}
	\label{tb:model9}
	\begin{center}
		\begin{tabular}{|c||c|c||c|c|c|c|c|c|c|c|c|}
			\hline\rm{model} \ref{tb:model9} & \multicolumn{11}{c|}{$U(4)\times U(2)_L\times U(2)_R\times USp(2)^2\times USp(4) $}\\
			\hline \hline			\rm{stack} & $N$ & $(n^1,l^1)\times(n^2,l^2)\times(n^3,l^3)$ & $n_{\Ysymm}$& $n_{\Yasymm}$ & $b$ & $b'$ & $c$ & $c'$ & 1 & 2 & 4\\
			\hline
			$a$ & 8 & $(1,-1)\times (-1,1)\times (1,-1)$ & 0 & 4  & 0 & 3 & 0 & -3 & -1 & 1 & 1\\
			$b$ & 4 & $(0,1)\times (-2,1)\times (-1,1)$ & -1 & 1  & - & - & 0 & -2 & -1 & 2 & 0\\
			$c$ & 4 & $(-1,0)\times (4,1)\times (-1,1)$ & 3 & -3  & - & - & - & - & 0 & 0 & -4\\
			\hline
			1 & 4 & $(1, 0)\times (1, 0)\times (2, 0)$& \multicolumn{9}{c|}{$x_A = 2x_B = \frac{5}{2}x_C = 10x_D$}\\
			2 & 2 & $(1, 0)\times (0, -1)\times (0, 2)$& \multicolumn{9}{c|}{$\beta^g_1=-3$, $\beta^g_2=-2$, $\beta^g_4=0$}\\
			4 & 2 & $(0, -1)\times (0, 1)\times (2, 0)$& \multicolumn{9}{c|}{$\chi_1=\frac{5}{\sqrt{2}}$, $\chi_2=2 \sqrt{2}$, $\chi_3=\sqrt{2}$}\\
			\hline
		\end{tabular}
	\end{center}
\end{table}
\begin{table}\scriptsize
	\caption{D6-brane configurations and intersection numbers of Model \ref{tb:model10}, and its MSSM gauge coupling relation is $g^2_a=g^2_b=\frac{1}{3}g^2_c=\frac{5}{11}(\frac{5}{3}g^2_Y)=\frac{16 \pi  e^{\phi ^4}}{5 \sqrt{3}}$.}
	\label{tb:model10}
	\begin{center}
		\begin{tabular}{|c||c|c||c|c|c|c|c|c|c|c|}
			\hline\rm{model} \ref{tb:model10} & \multicolumn{10}{c|}{$U(4)\times U(2)_L\times U(2)_R\times USp(2)\times USp(4) $}\\
			\hline \hline			\rm{stack} & $N$ & $(n^1,l^1)\times(n^2,l^2)\times(n^3,l^3)$ & $n_{\Ysymm}$& $n_{\Yasymm}$ & $b$ & $b'$ & $c$ & $c'$ & 1 & 3\\
			\hline
			$a$ & 8 & $(1,1)\times (-1,0)\times (-1,1)$ & 0 & 0  & 3 & 0 & 0 & -3 & 0 & 0\\
			$b$ & 4 & $(1,0)\times (1,-3)\times (1,1)$ & 2 & -2  & - & - & 4 & 0 & 0 & -3\\
			$c$ & 4 & $(-1,4)\times (0,1)\times (-1,1)$ & 3 & -3  & - & - & - & - & -4 & 1\\
			\hline
			1 & 2 & $(1, 0)\times (1, 0)\times (2, 0)$& \multicolumn{8}{c|}{$x_A = \frac{1}{12}x_B = \frac{1}{4}x_C = \frac{1}{12}x_D$}\\
			3 & 4 & $(0, -1)\times (1, 0)\times (0, 2)$& \multicolumn{8}{c|}{$\beta^g_1=-2$, $\beta^g_3=-2$}\\
			& & & \multicolumn{8}{c|}{$\chi_1=\frac{1}{2}$, $\chi_2=\frac{1}{6}$, $\chi_3=1$}\\
			\hline
		\end{tabular}
	\end{center}
\end{table}
\begin{table}\scriptsize
	\caption{D6-brane configurations and intersection numbers of Model \ref{tb:model11}, and its MSSM gauge coupling relation is $g^2_a=\frac{35}{66}g^2_b=\frac{7}{6}g^2_c=\frac{35}{32}(\frac{5}{3}g^2_Y)=\frac{8 \sqrt[4]{2} 5^{3/4} \pi  e^{\phi ^4}}{11 \sqrt{3}}$.}
	\label{tb:model11}
	\begin{center}
		\begin{tabular}{|c||c|c||c|c|c|c|c|c|c|}
			\hline\rm{model} \ref{tb:model11} & \multicolumn{9}{c|}{$U(4)\times U(2)_L\times U(2)_R\times USp(2) $}\\
			\hline \hline			\rm{stack} & $N$ & $(n^1,l^1)\times(n^2,l^2)\times(n^3,l^3)$ & $n_{\Ysymm}$& $n_{\Yasymm}$ & $b$ & $b'$ & $c$ & $c'$ & 3\\
			\hline
			$a$ & 8 & $(1,-1)\times (1,0)\times (1,1)$ & 0 & 0  & -3 & 6 & 0 & -3 & 0\\
			$b$ & 4 & $(-2,1)\times (0,1)\times (-5,1)$ & -9 & 9  & - & - & 0 & 8 & 10\\
			$c$ & 4 & $(-2,1)\times (-1,1)\times (1,1)$ & -2 & -6  & - & - & - & - & -2\\
			\hline
			3 & 2 & $(0, -1)\times (1, 0)\times (0, 2)$& \multicolumn{7}{c|}{$x_A = \frac{5}{6}x_B = 10x_C = \frac{5}{6}x_D$}\\
			& & & \multicolumn{7}{c|}{$\beta^g_3=6$}\\
			& & & \multicolumn{7}{c|}{$\chi_1=\sqrt{10}$, $\chi_2=\frac{\sqrt{\frac{5}{2}}}{6}$, $\chi_3=2 \sqrt{10}$}\\
			\hline
		\end{tabular}
	\end{center}
\end{table}
\begin{table}\scriptsize
	\caption{D6-brane configurations and intersection numbers of Model \ref{tb:model12}, and its MSSM gauge coupling relation is $g^2_a=\frac{4}{9}g^2_b=\frac{10}{9}g^2_c=\frac{50}{47}(\frac{5}{3}g^2_Y)=\frac{16}{27} 2^{3/4} \sqrt{5} \pi  e^{\phi ^4}$.}
	\label{tb:model12}
	\begin{center}
		\begin{tabular}{|c||c|c||c|c|c|c|c|c|c|c|c|}
			\hline\rm{model} \ref{tb:model12} & \multicolumn{11}{c|}{$U(4)\times U(2)_L\times U(2)_R\times USp(2)^2\times USp(4) $}\\
			\hline \hline			\rm{stack} & $N$ & $(n^1,l^1)\times(n^2,l^2)\times(n^3,l^3)$ & $n_{\Ysymm}$& $n_{\Yasymm}$ & $b$ & $b'$ & $c$ & $c'$ & 1 & 3 & 4\\
			\hline
			$a$ & 8 & $(-1,-1)\times (1,1)\times (1,1)$ & 0 & -4  & 0 & 3 & 0 & -3 & 1 & -1 & -1\\
			$b$ & 4 & $(-4,1)\times (-1,0)\times (1,1)$ & -3 & 3  & - & - & 0 & 2 & 0 & 0 & 4\\
			$c$ & 4 & $(-2,-1)\times (0,1)\times (1,1)$ & 1 & -1  & - & - & - & - & 1 & -2 & 0\\
			\hline
			1 & 4 & $(1, 0)\times (1, 0)\times (2, 0)$& \multicolumn{9}{c|}{$x_A = \frac{5}{2}x_B = 2x_C = 10x_D$}\\
			3 & 2 & $(0, -1)\times (1, 0)\times (0, 2)$& \multicolumn{9}{c|}{$\beta^g_1=-3$, $\beta^g_3=-2$, $\beta^g_4=0$}\\
			4 & 2 & $(0, -1)\times (0, 1)\times (2, 0)$& \multicolumn{9}{c|}{$\chi_1=2 \sqrt{2}$, $\chi_2=\frac{5}{\sqrt{2}}$, $\chi_3=\sqrt{2}$}\\
			\hline
		\end{tabular}
	\end{center}
\end{table}
\begin{table}\scriptsize
	\caption{D6-brane configurations and intersection numbers of Model \ref{tb:model13}, and its MSSM gauge coupling relation is $g^2_a=2g^2_b=\frac{2}{3}g^2_c=\frac{10}{13}(\frac{5}{3}g^2_Y)=\frac{16}{5} \sqrt{\frac{2}{3}} \pi  e^{\phi ^4}$.}
	\label{tb:model13}
	\begin{center}
		\begin{tabular}{|c||c|c||c|c|c|c|c|c|c|}
			\hline\rm{model} \ref{tb:model13} & \multicolumn{9}{c|}{$U(4)\times U(2)_L\times U(2)_R\times USp(4) $}\\
			\hline \hline			\rm{stack} & $N$ & $(n^1,l^1)\times(n^2,l^2)\times(n^3,l^3)$ & $n_{\Ysymm}$& $n_{\Yasymm}$ & $b$ & $b'$ & $c$ & $c'$ & 3\\
			\hline
			$a$ & 8 & $(1,1)\times (-1,0)\times (-1,1)$ & 0 & 0  & 3 & 0 & 0 & -3 & 0\\
			$b$ & 4 & $(1,0)\times (2,-3)\times (1,1)$ & 1 & -1  & - & - & 8 & 0 & -3\\
			$c$ & 4 & $(-1,4)\times (0,1)\times (-1,1)$ & 3 & -3  & - & - & - & - & 1\\
			\hline
			3 & 4 & $(0, -1)\times (1, 0)\times (0, 2)$& \multicolumn{7}{c|}{$x_A = \frac{1}{6}x_B = \frac{1}{4}x_C = \frac{1}{6}x_D$}\\
			& & & \multicolumn{7}{c|}{$\beta^g_3=-2$}\\
			& & & \multicolumn{7}{c|}{$\chi_1=\frac{1}{2}$, $\chi_2=\frac{1}{3}$, $\chi_3=1$}\\
			\hline
		\end{tabular}
	\end{center}
\end{table}
\begin{table}\scriptsize
	\caption{D6-brane configurations and intersection numbers of Model \ref{tb:model14}, and its MSSM gauge coupling relation is $g^2_a=g^2_b=2g^2_c=\frac{10}{7}(\frac{5}{3}g^2_Y)=\frac{16 \sqrt[4]{2} \pi  e^{\phi ^4}}{3 \sqrt{3}}$.}
	\label{tb:model14}
	\begin{center}
		\begin{tabular}{|c||c|c||c|c|c|c|c|c|c|c|c|}
			\hline\rm{model} \ref{tb:model14} & \multicolumn{11}{c|}{$U(4)\times U(2)_L\times U(2)_R\times USp(2)^3 $}\\
			\hline \hline			\rm{stack} & $N$ & $(n^1,l^1)\times(n^2,l^2)\times(n^3,l^3)$ & $n_{\Ysymm}$& $n_{\Yasymm}$ & $b$ & $b'$ & $c$ & $c'$ & 1 & 2 & 4\\
			\hline
			$a$ & 8 & $(1,1)\times (1,0)\times (1,-1)$ & 0 & 0  & 3 & 0 & 0 & -3 & 0 & 1 & -1\\
			$b$ & 4 & $(-1,0)\times (-1,3)\times (1,1)$ & 2 & -2  & - & - & -3 & 0 & 0 & 0 & 1\\
			$c$ & 4 & $(0,1)\times (-2,3)\times (-1,1)$ & 1 & -1  & - & - & - & - & -3 & 2 & 0\\
			\hline
			1 & 2 & $(1, 0)\times (1, 0)\times (2, 0)$& \multicolumn{9}{c|}{$x_A = \frac{2}{3}x_B = 2x_C = \frac{2}{3}x_D$}\\
			2 & 2 & $(1, 0)\times (0, -1)\times (0, 2)$& \multicolumn{9}{c|}{$\beta^g_1=-3$, $\beta^g_2=-2$, $\beta^g_4=-3$}\\
			4 & 2 & $(0, -1)\times (0, 1)\times (2, 0)$& \multicolumn{9}{c|}{$\chi_1=\sqrt{2}$, $\chi_2=\frac{\sqrt{2}}{3}$, $\chi_3=2 \sqrt{2}$}\\
			\hline
		\end{tabular}
	\end{center}
\end{table}
\begin{table}\scriptsize
	\caption{D6-brane configurations and intersection numbers of Model \ref{tb:model15}, and its MSSM gauge coupling relation is $g^2_a=2g^2_b=2g^2_c=\frac{10}{7}(\frac{5}{3}g^2_Y)=\frac{8 \pi  e^{\phi ^4}}{\sqrt{3}}$.}
	\label{tb:model15}
	\begin{center}
		\begin{tabular}{|c||c|c||c|c|c|c|c|c|c|c|}
			\hline\rm{model} \ref{tb:model15} & \multicolumn{10}{c|}{$U(4)\times U(2)_L\times U(2)_R\times USp(2)^2 $}\\
			\hline \hline			\rm{stack} & $N$ & $(n^1,l^1)\times(n^2,l^2)\times(n^3,l^3)$ & $n_{\Ysymm}$& $n_{\Yasymm}$ & $b$ & $b'$ & $c$ & $c'$ & 2 & 4\\
			\hline
			$a$ & 8 & $(1,1)\times (1,0)\times (1,-1)$ & 0 & 0  & 3 & 0 & 0 & -3 & 1 & -1\\
			$b$ & 4 & $(-1,0)\times (-2,3)\times (1,1)$ & 1 & -1  & - & - & 0 & 0 & 0 & 2\\
			$c$ & 4 & $(0,1)\times (-2,3)\times (-1,1)$ & 1 & -1  & - & - & - & - & 2 & 0\\
			\hline
			2 & 2 & $(1, 0)\times (0, -1)\times (0, 2)$& \multicolumn{8}{c|}{$x_A = \frac{2}{3}x_B = x_C = \frac{2}{3}x_D$}\\
			4 & 2 & $(0, -1)\times (0, 1)\times (2, 0)$& \multicolumn{8}{c|}{$\beta^g_2=-2$, $\beta^g_4=-2$}\\
			& & & \multicolumn{8}{c|}{$\chi_1=1$, $\chi_2=\frac{2}{3}$, $\chi_3=2$}\\
			\hline
		\end{tabular}
	\end{center}
\end{table}
\begin{table}\scriptsize
	\caption{D6-brane configurations and intersection numbers of Model \ref{tb:model16}, and its MSSM gauge coupling relation is $g^2_a=\frac{54}{19}g^2_b=g^2_c=(\frac{5}{3}g^2_Y)=\frac{48}{19} \sqrt[4]{2} \pi  e^{\phi ^4}$.}
	\label{tb:model16}
	\begin{center}
		\begin{tabular}{|c||c|c||c|c|c|c|c|c|c|c|c|}
			\hline\rm{model} \ref{tb:model16} & \multicolumn{11}{c|}{$U(4)\times U(2)_L\times U(2)_R\times USp(2)^3 $}\\
			\hline \hline			\rm{stack} & $N$ & $(n^1,l^1)\times(n^2,l^2)\times(n^3,l^3)$ & $n_{\Ysymm}$& $n_{\Yasymm}$ & $b$ & $b'$ & $c$ & $c'$ & 2 & 3 & 4\\
			\hline
			$a$ & 8 & $(1,1)\times (1,0)\times (1,-1)$ & 0 & 0  & 2 & 1 & 0 & -3 & 1 & 0 & -1\\
			$b$ & 4 & $(-1,0)\times (-2,1)\times (1,3)$ & -5 & 5  & - & - & 10 & 7 & 0 & -1 & 6\\
			$c$ & 4 & $(0,1)\times (-1,3)\times (-1,1)$ & 2 & -2  & - & - & - & - & 1 & 0 & 0\\
			\hline
			2 & 2 & $(1, 0)\times (0, -1)\times (0, 2)$& \multicolumn{9}{c|}{$x_A = \frac{1}{3}x_B = \frac{1}{18}x_C = \frac{1}{3}x_D$}\\
			3 & 2 & $(0, -1)\times (1, 0)\times (0, 2)$& \multicolumn{9}{c|}{$\beta^g_2=-3$, $\beta^g_3=-5$, $\beta^g_4=2$}\\
			4 & 2 & $(0, -1)\times (0, 1)\times (2, 0)$& \multicolumn{9}{c|}{$\chi_1=\frac{1}{3 \sqrt{2}}$, $\chi_2=\sqrt{2}$, $\chi_3=\frac{\sqrt{2}}{3}$}\\
			\hline
		\end{tabular}
	\end{center}
\end{table}
\begin{table}\scriptsize
	\caption{D6-brane configurations and intersection numbers of Model \ref{tb:model17}, and its MSSM gauge coupling relation is $g^2_a=\frac{5}{6}g^2_b=\frac{11}{6}g^2_c=\frac{11}{8}(\frac{5}{3}g^2_Y)=\frac{8 \sqrt[4]{2} 5^{3/4} \pi  e^{\phi ^4}}{7 \sqrt{3}}$.}
	\label{tb:model17}
	\begin{center}
		\begin{tabular}{|c||c|c||c|c|c|c|c|c|c|}
			\hline\rm{model} \ref{tb:model17} & \multicolumn{9}{c|}{$U(4)\times U(2)_L\times U(2)_R\times USp(2) $}\\
			\hline \hline			\rm{stack} & $N$ & $(n^1,l^1)\times(n^2,l^2)\times(n^3,l^3)$ & $n_{\Ysymm}$& $n_{\Yasymm}$ & $b$ & $b'$ & $c$ & $c'$ & 3\\
			\hline
			$a$ & 8 & $(1,-1)\times (1,0)\times (1,1)$ & 0 & 0  & 3 & 0 & 0 & -3 & 0\\
			$b$ & 4 & $(-2,5)\times (0,1)\times (-1,1)$ & 3 & -3  & - & - & -8 & 0 & 2\\
			$c$ & 4 & $(-2,1)\times (-1,1)\times (1,1)$ & -2 & -6  & - & - & - & - & -2\\
			\hline
			3 & 2 & $(0, -1)\times (1, 0)\times (0, 2)$& \multicolumn{7}{c|}{$x_A = \frac{1}{6}x_B = \frac{2}{5}x_C = \frac{1}{6}x_D$}\\
			& & & \multicolumn{7}{c|}{$\beta^g_3=-2$}\\
			& & & \multicolumn{7}{c|}{$\chi_1=\sqrt{\frac{2}{5}}$, $\chi_2=\frac{\sqrt{\frac{5}{2}}}{6}$, $\chi_3=2 \sqrt{\frac{2}{5}}$}\\
			\hline
		\end{tabular}
	\end{center}
\end{table}
\begin{table}\scriptsize
	\caption{D6-brane configurations and intersection numbers of Model \ref{tb:model18}, and its MSSM gauge coupling relation is $g^2_a=\frac{7}{6}g^2_b=\frac{1}{6}g^2_c=\frac{1}{4}(\frac{5}{3}g^2_Y)=\frac{16}{135} 26^{3/4} \pi  e^{\phi ^4}$.}
	\label{tb:model18}
	\begin{center}
		\begin{tabular}{|c||c|c||c|c|c|c|c|c|c|c|}
			\hline\rm{model} \ref{tb:model18} & \multicolumn{10}{c|}{$U(4)\times U(2)_L\times U(2)_R\times USp(2)\times USp(4) $}\\
			\hline \hline			\rm{stack} & $N$ & $(n^1,l^1)\times(n^2,l^2)\times(n^3,l^3)$ & $n_{\Ysymm}$& $n_{\Yasymm}$ & $b$ & $b'$ & $c$ & $c'$ & 1 & 4\\
			\hline
			$a$ & 8 & $(1,-1)\times (-2,1)\times (1,-1)$ & 0 & 8  & 0 & 3 & 0 & -3 & -1 & 2\\
			$b$ & 4 & $(-2,1)\times (1,1)\times (-1,1)$ & -2 & -6  & - & - & 0 & -2 & -1 & -2\\
			$c$ & 4 & $(-4,-1)\times (1,0)\times (-1,1)$ & 3 & -3  & - & - & - & - & 0 & -4\\
			\hline
			1 & 4 & $(1, 0)\times (1, 0)\times (2, 0)$& \multicolumn{8}{c|}{$x_A = \frac{13}{2}x_B = \frac{13}{8}x_C = 26x_D$}\\
			4 & 2 & $(0, -1)\times (0, 1)\times (2, 0)$& \multicolumn{8}{c|}{$\beta^g_1=-3$, $\beta^g_4=4$}\\
			& & & \multicolumn{8}{c|}{$\chi_1=\sqrt{\frac{13}{2}}$, $\chi_2=2 \sqrt{26}$, $\chi_3=\frac{\sqrt{\frac{13}{2}}}{2}$}\\
			\hline
		\end{tabular}
	\end{center}
\end{table}
\begin{table}\scriptsize
	\caption{D6-brane configurations and intersection numbers of Model \ref{tb:model19}, and its MSSM gauge coupling relation is $g^2_a=\frac{17}{9}g^2_b=\frac{4}{9}g^2_c=\frac{4}{7}(\frac{5}{3}g^2_Y)=\frac{32 \pi  e^{\phi ^4}}{15}$.}
	\label{tb:model19}
	\begin{center}
		\begin{tabular}{|c||c|c||c|c|c|c|c|c|c|c|c|}
			\hline\rm{model} \ref{tb:model19} & \multicolumn{11}{c|}{$U(4)\times U(2)_L\times U(2)_R\times USp(2)^3 $}\\
			\hline \hline			\rm{stack} & $N$ & $(n^1,l^1)\times(n^2,l^2)\times(n^3,l^3)$ & $n_{\Ysymm}$& $n_{\Yasymm}$ & $b$ & $b'$ & $c$ & $c'$ & 2 & 3 & 4\\
			\hline
			$a$ & 8 & $(1,1)\times (1,0)\times (1,-1)$ & 0 & 0  & 3 & 0 & 0 & -3 & 1 & 0 & -1\\
			$b$ & 4 & $(2,-1)\times (1,-1)\times (1,1)$ & -2 & -6  & - & - & 7 & 0 & -1 & -2 & 2\\
			$c$ & 4 & $(-1,4)\times (0,1)\times (-1,1)$ & 3 & -3  & - & - & - & - & 0 & 1 & 0\\
			\hline
			2 & 2 & $(1, 0)\times (0, -1)\times (0, 2)$& \multicolumn{9}{c|}{$x_A = \frac{1}{9}x_B = \frac{1}{4}x_C = \frac{1}{9}x_D$}\\
			3 & 2 & $(0, -1)\times (1, 0)\times (0, 2)$& \multicolumn{9}{c|}{$\beta^g_2=-3$, $\beta^g_3=-3$, $\beta^g_4=-2$}\\
			4 & 2 & $(0, -1)\times (0, 1)\times (2, 0)$& \multicolumn{9}{c|}{$\chi_1=\frac{1}{2}$, $\chi_2=\frac{2}{9}$, $\chi_3=1$}\\
			\hline
		\end{tabular}
	\end{center}
\end{table}
\begin{table}\scriptsize
	\caption{D6-brane configurations and intersection numbers of Model \ref{tb:model20}, and its MSSM gauge coupling relation is $g^2_a=\frac{1}{3}g^2_b=g^2_c=(\frac{5}{3}g^2_Y)=\frac{16 \pi  e^{\phi ^4}}{5 \sqrt{3}}$.}
	\label{tb:model20}
	\begin{center}
		\begin{tabular}{|c||c|c||c|c|c|c|c|c|c|c|}
			\hline\rm{model} \ref{tb:model20} & \multicolumn{10}{c|}{$U(4)\times U(2)_L\times U(2)_R\times USp(2)\times USp(4) $}\\
			\hline \hline			\rm{stack} & $N$ & $(n^1,l^1)\times(n^2,l^2)\times(n^3,l^3)$ & $n_{\Ysymm}$& $n_{\Yasymm}$ & $b$ & $b'$ & $c$ & $c'$ & 1 & 3\\
			\hline
			$a$ & 8 & $(-1,1)\times (-1,0)\times (1,1)$ & 0 & 0  & 3 & 0 & 0 & -3 & 0 & 0\\
			$b$ & 4 & $(-1,4)\times (0,1)\times (-1,1)$ & 3 & -3  & - & - & -4 & 0 & -4 & 1\\
			$c$ & 4 & $(1,0)\times (1,-3)\times (1,1)$ & 2 & -2  & - & - & - & - & 0 & -3\\
			\hline
			1 & 2 & $(1, 0)\times (1, 0)\times (2, 0)$& \multicolumn{8}{c|}{$x_A = \frac{1}{12}x_B = \frac{1}{4}x_C = \frac{1}{12}x_D$}\\
			3 & 4 & $(0, -1)\times (1, 0)\times (0, 2)$& \multicolumn{8}{c|}{$\beta^g_1=-2$, $\beta^g_3=-2$}\\
			& & & \multicolumn{8}{c|}{$\chi_1=\frac{1}{2}$, $\chi_2=\frac{1}{6}$, $\chi_3=1$}\\
			\hline
		\end{tabular}
	\end{center}
\end{table}
\begin{table}\scriptsize
	\caption{D6-brane configurations and intersection numbers of Model \ref{tb:model21}, and its MSSM gauge coupling relation is $g^2_a=\frac{13}{6}g^2_b=\frac{1}{2}g^2_c=\frac{5}{8}(\frac{5}{3}g^2_Y)=\frac{16}{5} \sqrt{\frac{2}{3}} \pi  e^{\phi ^4}$.}
	\label{tb:model21}
	\begin{center}
		\begin{tabular}{|c||c|c||c|c|c|c|c|c|c|}
			\hline\rm{model} \ref{tb:model21} & \multicolumn{9}{c|}{$U(4)\times U(2)_L\times U(2)_R\times USp(4) $}\\
			\hline \hline			\rm{stack} & $N$ & $(n^1,l^1)\times(n^2,l^2)\times(n^3,l^3)$ & $n_{\Ysymm}$& $n_{\Yasymm}$ & $b$ & $b'$ & $c$ & $c'$ & 4\\
			\hline
			$a$ & 8 & $(-2,1)\times (-1,0)\times (1,1)$ & -1 & 1  & 3 & 0 & 0 & -3 & 2\\
			$b$ & 4 & $(-1,2)\times (-1,1)\times (-1,1)$ & 2 & 6  & - & - & 4 & 0 & 1\\
			$c$ & 4 & $(1,0)\times (1,-3)\times (1,1)$ & 2 & -2  & - & - & - & - & 1\\
			\hline
			4 & 4 & $(0, -1)\times (0, 1)\times (2, 0)$& \multicolumn{7}{c|}{$x_A = \frac{4}{3}x_B = 8x_C = \frac{8}{3}x_D$}\\
			& & & \multicolumn{7}{c|}{$\beta^g_4=0$}\\
			& & & \multicolumn{7}{c|}{$\chi_1=4$, $\chi_2=\frac{2}{3}$, $\chi_3=4$}\\
			\hline
		\end{tabular}
	\end{center}
\end{table}
\begin{table}\scriptsize
	\caption{D6-brane configurations and intersection numbers of Model \ref{tb:model22}, and its MSSM gauge coupling relation is $g^2_a=\frac{27}{11}g^2_b=2g^2_c=\frac{10}{7}(\frac{5}{3}g^2_Y)=\frac{48}{11} \sqrt[4]{2} \pi  e^{\phi ^4}$.}
	\label{tb:model22}
	\begin{center}
		\begin{tabular}{|c||c|c||c|c|c|c|c|c|c|c|c|}
			\hline\rm{model} \ref{tb:model22} & \multicolumn{11}{c|}{$U(4)\times U(2)_L\times U(2)_R\times USp(2)^3 $}\\
			\hline \hline			\rm{stack} & $N$ & $(n^1,l^1)\times(n^2,l^2)\times(n^3,l^3)$ & $n_{\Ysymm}$& $n_{\Yasymm}$ & $b$ & $b'$ & $c$ & $c'$ & 1 & 2 & 4\\
			\hline
			$a$ & 8 & $(1,1)\times (1,0)\times (1,-1)$ & 0 & 0  & 2 & 1 & 0 & -3 & 0 & 1 & -1\\
			$b$ & 4 & $(-1,0)\times (-1,1)\times (1,3)$ & -2 & 2  & - & - & 2 & 5 & 0 & 0 & 3\\
			$c$ & 4 & $(0,1)\times (-2,3)\times (-1,1)$ & 1 & -1  & - & - & - & - & -3 & 2 & 0\\
			\hline
			1 & 2 & $(1, 0)\times (1, 0)\times (2, 0)$& \multicolumn{9}{c|}{$x_A = \frac{2}{3}x_B = \frac{2}{9}x_C = \frac{2}{3}x_D$}\\
			2 & 2 & $(1, 0)\times (0, -1)\times (0, 2)$& \multicolumn{9}{c|}{$\beta^g_1=-3$, $\beta^g_2=-2$, $\beta^g_4=-1$}\\
			4 & 2 & $(0, -1)\times (0, 1)\times (2, 0)$& \multicolumn{9}{c|}{$\chi_1=\frac{\sqrt{2}}{3}$, $\chi_2=\sqrt{2}$, $\chi_3=\frac{2 \sqrt{2}}{3}$}\\
			\hline
		\end{tabular}
	\end{center}
\end{table}
\begin{table}\scriptsize
	\caption{D6-brane configurations and intersection numbers of Model \ref{tb:model23}, and its MSSM gauge coupling relation is $g^2_a=\frac{11}{6}g^2_b=\frac{1}{6}g^2_c=\frac{1}{4}(\frac{5}{3}g^2_Y)=\frac{8}{405} \sqrt{2} 161^{3/4} \pi  e^{\phi ^4}$.}
	\label{tb:model23}
	\begin{center}
		\begin{tabular}{|c||c|c||c|c|c|c|c|c|c|c|}
			\hline\rm{model} \ref{tb:model23} & \multicolumn{10}{c|}{$U(4)\times U(2)_L\times U(2)_R\times USp(2)^2 $}\\
			\hline \hline			\rm{stack} & $N$ & $(n^1,l^1)\times(n^2,l^2)\times(n^3,l^3)$ & $n_{\Ysymm}$& $n_{\Yasymm}$ & $b$ & $b'$ & $c$ & $c'$ & 1 & 4\\
			\hline
			$a$ & 8 & $(-1,-1)\times (1,1)\times (1,1)$ & 0 & -4  & 3 & 0 & 0 & -3 & 1 & -1\\
			$b$ & 4 & $(-2,1)\times (2,1)\times (-1,1)$ & -3 & -13  & - & - & 7 & 0 & -1 & -4\\
			$c$ & 4 & $(1,-4)\times (1,0)\times (1,1)$ & 3 & -3  & - & - & - & - & 0 & 1\\
			\hline
			1 & 2 & $(1, 0)\times (1, 0)\times (2, 0)$& \multicolumn{8}{c|}{$x_A = 28x_B = \frac{28}{23}x_C = 7x_D$}\\
			4 & 2 & $(0, -1)\times (0, 1)\times (2, 0)$& \multicolumn{8}{c|}{$\beta^g_1=-3$, $\beta^g_4=1$}\\
			& & & \multicolumn{8}{c|}{$\chi_1=\sqrt{\frac{7}{23}}$, $\chi_2=\sqrt{161}$, $\chi_3=8 \sqrt{\frac{7}{23}}$}\\
			\hline
		\end{tabular}
	\end{center}
\end{table}
\begin{table}\scriptsize
	\caption{D6-brane configurations and intersection numbers of Model \ref{tb:model24}, and its MSSM gauge coupling relation is $g^2_a=\frac{2}{3}g^2_b=2g^2_c=\frac{10}{7}(\frac{5}{3}g^2_Y)=\frac{16}{5} \sqrt{\frac{2}{3}} \pi  e^{\phi ^4}$.}
	\label{tb:model24}
	\begin{center}
		\begin{tabular}{|c||c|c||c|c|c|c|c|c|c|}
			\hline\rm{model} \ref{tb:model24} & \multicolumn{9}{c|}{$U(4)\times U(2)_L\times U(2)_R\times USp(4) $}\\
			\hline \hline			\rm{stack} & $N$ & $(n^1,l^1)\times(n^2,l^2)\times(n^3,l^3)$ & $n_{\Ysymm}$& $n_{\Yasymm}$ & $b$ & $b'$ & $c$ & $c'$ & 3\\
			\hline
			$a$ & 8 & $(-1,1)\times (-1,0)\times (1,1)$ & 0 & 0  & 3 & 0 & 0 & -3 & 0\\
			$b$ & 4 & $(-1,4)\times (0,1)\times (-1,1)$ & 3 & -3  & - & - & -8 & 0 & 1\\
			$c$ & 4 & $(1,0)\times (2,-3)\times (1,1)$ & 1 & -1  & - & - & - & - & -3\\
			\hline
			3 & 4 & $(0, -1)\times (1, 0)\times (0, 2)$& \multicolumn{7}{c|}{$x_A = \frac{1}{6}x_B = \frac{1}{4}x_C = \frac{1}{6}x_D$}\\
			& & & \multicolumn{7}{c|}{$\beta^g_3=-2$}\\
			& & & \multicolumn{7}{c|}{$\chi_1=\frac{1}{2}$, $\chi_2=\frac{1}{3}$, $\chi_3=1$}\\
			\hline
		\end{tabular}
	\end{center}
\end{table}
\begin{table}\scriptsize
	\caption{D6-brane configurations and intersection numbers of Model \ref{tb:model25}, and its MSSM gauge coupling relation is $g^2_a=\frac{13}{5}g^2_b=3g^2_c=\frac{5}{3}(\frac{5}{3}g^2_Y)=\frac{16}{5} \sqrt{3} \pi  e^{\phi ^4}$.}
	\label{tb:model25}
	\begin{center}
		\begin{tabular}{|c||c|c||c|c|c|c|c|c|c|c|}
			\hline\rm{model} \ref{tb:model25} & \multicolumn{10}{c|}{$U(4)\times U(2)_L\times U(2)_R\times USp(2)\times USp(4) $}\\
			\hline \hline			\rm{stack} & $N$ & $(n^1,l^1)\times(n^2,l^2)\times(n^3,l^3)$ & $n_{\Ysymm}$& $n_{\Yasymm}$ & $b$ & $b'$ & $c$ & $c'$ & 1 & 3\\
			\hline
			$a$ & 8 & $(1,1)\times (-1,0)\times (-1,1)$ & 0 & 0  & 2 & 1 & 0 & -3 & 0 & 0\\
			$b$ & 4 & $(1,0)\times (1,-1)\times (1,3)$ & -2 & 2  & - & - & 8 & 4 & 0 & -1\\
			$c$ & 4 & $(-1,4)\times (0,1)\times (-1,1)$ & 3 & -3  & - & - & - & - & -4 & 1\\
			\hline
			1 & 2 & $(1, 0)\times (1, 0)\times (2, 0)$& \multicolumn{8}{c|}{$x_A = \frac{3}{4}x_B = \frac{1}{4}x_C = \frac{3}{4}x_D$}\\
			3 & 4 & $(0, -1)\times (1, 0)\times (0, 2)$& \multicolumn{8}{c|}{$\beta^g_1=-2$, $\beta^g_3=-4$}\\
			& & & \multicolumn{8}{c|}{$\chi_1=\frac{1}{2}$, $\chi_2=\frac{3}{2}$, $\chi_3=1$}\\
			\hline
		\end{tabular}
	\end{center}
\end{table}
\begin{table}\scriptsize
	\caption{D6-brane configurations and intersection numbers of Model \ref{tb:model26}, and its MSSM gauge coupling relation is $g^2_a=\frac{18}{5}g^2_b=2g^2_c=\frac{10}{7}(\frac{5}{3}g^2_Y)=\frac{24 \pi  e^{\phi ^4}}{5}$.}
	\label{tb:model26}
	\begin{center}
		\begin{tabular}{|c||c|c||c|c|c|c|c|c|c|c|}
			\hline\rm{model} \ref{tb:model26} & \multicolumn{10}{c|}{$U(4)\times U(2)_L\times U(2)_R\times USp(2)^2 $}\\
			\hline \hline			\rm{stack} & $N$ & $(n^1,l^1)\times(n^2,l^2)\times(n^3,l^3)$ & $n_{\Ysymm}$& $n_{\Yasymm}$ & $b$ & $b'$ & $c$ & $c'$ & 2 & 4\\
			\hline
			$a$ & 8 & $(1,1)\times (1,0)\times (1,-1)$ & 0 & 0  & 2 & 1 & 0 & -3 & 1 & -1\\
			$b$ & 4 & $(-1,0)\times (-2,1)\times (1,3)$ & -5 & 5  & - & - & 8 & 8 & 0 & 6\\
			$c$ & 4 & $(0,1)\times (-2,3)\times (-1,1)$ & 1 & -1  & - & - & - & - & 2 & 0\\
			\hline
			2 & 2 & $(1, 0)\times (0, -1)\times (0, 2)$& \multicolumn{8}{c|}{$x_A = \frac{2}{3}x_B = \frac{1}{9}x_C = \frac{2}{3}x_D$}\\
			4 & 2 & $(0, -1)\times (0, 1)\times (2, 0)$& \multicolumn{8}{c|}{$\beta^g_2=-2$, $\beta^g_4=2$}\\
			& & & \multicolumn{8}{c|}{$\chi_1=\frac{1}{3}$, $\chi_2=2$, $\chi_3=\frac{2}{3}$}\\
			\hline
		\end{tabular}
	\end{center}
\end{table}
\begin{table}\scriptsize
	\caption{D6-brane configurations and intersection numbers of Model \ref{tb:model27}, and its MSSM gauge coupling relation is $g^2_a=\frac{21}{10}g^2_b=\frac{7}{2}g^2_c=\frac{7}{4}(\frac{5}{3}g^2_Y)=\frac{8}{5} 6^{3/4} \pi  e^{\phi ^4}$.}
	\label{tb:model27}
	\begin{center}
		\begin{tabular}{|c||c|c||c|c|c|c|c|c|c|}
			\hline\rm{model} \ref{tb:model27} & \multicolumn{9}{c|}{$U(4)\times U(2)_L\times U(2)_R\times USp(4) $}\\
			\hline \hline			\rm{stack} & $N$ & $(n^1,l^1)\times(n^2,l^2)\times(n^3,l^3)$ & $n_{\Ysymm}$& $n_{\Yasymm}$ & $b$ & $b'$ & $c$ & $c'$ & 4\\
			\hline
			$a$ & 8 & $(2,1)\times (1,0)\times (1,-1)$ & 1 & -1  & 2 & 1 & 0 & -3 & -2\\
			$b$ & 4 & $(1,0)\times (1,-1)\times (1,3)$ & -2 & 2  & - & - & 0 & 4 & 3\\
			$c$ & 4 & $(-1,2)\times (-1,1)\times (-1,1)$ & 2 & 6  & - & - & - & - & 1\\
			\hline
			4 & 4 & $(0, -1)\times (0, 1)\times (2, 0)$& \multicolumn{7}{c|}{$x_A = 2x_B = \frac{4}{3}x_C = 4x_D$}\\
			& & & \multicolumn{7}{c|}{$\beta^g_4=2$}\\
			& & & \multicolumn{7}{c|}{$\chi_1=2 \sqrt{\frac{2}{3}}$, $\chi_2=\sqrt{6}$, $\chi_3=2 \sqrt{\frac{2}{3}}$}\\
			\hline
		\end{tabular}
	\end{center}
\end{table}
\begin{table}\scriptsize
	\caption{D6-brane configurations and intersection numbers of Model \ref{tb:model28}, and its MSSM gauge coupling relation is $g^2_a=\frac{4}{9}g^2_b=\frac{17}{9}g^2_c=\frac{85}{61}(\frac{5}{3}g^2_Y)=\frac{32 \pi  e^{\phi ^4}}{15}$.}
	\label{tb:model28}
	\begin{center}
		\begin{tabular}{|c||c|c||c|c|c|c|c|c|c|c|c|}
			\hline\rm{model} \ref{tb:model28} & \multicolumn{11}{c|}{$U(4)\times U(2)_L\times U(2)_R\times USp(2)^3 $}\\
			\hline \hline			\rm{stack} & $N$ & $(n^1,l^1)\times(n^2,l^2)\times(n^3,l^3)$ & $n_{\Ysymm}$& $n_{\Yasymm}$ & $b$ & $b'$ & $c$ & $c'$ & 2 & 3 & 4\\
			\hline
			$a$ & 8 & $(1,-1)\times (1,0)\times (1,1)$ & 0 & 0  & 3 & 0 & 0 & -3 & -1 & 0 & 1\\
			$b$ & 4 & $(-1,4)\times (0,1)\times (-1,1)$ & 3 & -3  & - & - & -7 & 0 & 0 & 1 & 0\\
			$c$ & 4 & $(-2,1)\times (-1,1)\times (1,1)$ & -2 & -6  & - & - & - & - & -1 & -2 & 2\\
			\hline
			2 & 2 & $(1, 0)\times (0, -1)\times (0, 2)$& \multicolumn{9}{c|}{$x_A = \frac{1}{9}x_B = \frac{1}{4}x_C = \frac{1}{9}x_D$}\\
			3 & 2 & $(0, -1)\times (1, 0)\times (0, 2)$& \multicolumn{9}{c|}{$\beta^g_2=-3$, $\beta^g_3=-3$, $\beta^g_4=-2$}\\
			4 & 2 & $(0, -1)\times (0, 1)\times (2, 0)$& \multicolumn{9}{c|}{$\chi_1=\frac{1}{2}$, $\chi_2=\frac{2}{9}$, $\chi_3=1$}\\
			\hline
		\end{tabular}
	\end{center}
\end{table}
\begin{table}\scriptsize
	\caption{D6-brane configurations and intersection numbers of Model \ref{tb:model29}, and its MSSM gauge coupling relation is $g^2_a=\frac{1}{2}g^2_b=\frac{13}{6}g^2_c=\frac{65}{44}(\frac{5}{3}g^2_Y)=\frac{16}{5} \sqrt{\frac{2}{3}} \pi  e^{\phi ^4}$.}
	\label{tb:model29}
	\begin{center}
		\begin{tabular}{|c||c|c||c|c|c|c|c|c|c|}
			\hline\rm{model} \ref{tb:model29} & \multicolumn{9}{c|}{$U(4)\times U(2)_L\times U(2)_R\times USp(4) $}\\
			\hline \hline			\rm{stack} & $N$ & $(n^1,l^1)\times(n^2,l^2)\times(n^3,l^3)$ & $n_{\Ysymm}$& $n_{\Yasymm}$ & $b$ & $b'$ & $c$ & $c'$ & 4\\
			\hline
			$a$ & 8 & $(2,1)\times (1,0)\times (1,-1)$ & 1 & -1  & 3 & 0 & 0 & -3 & -2\\
			$b$ & 4 & $(1,0)\times (1,-3)\times (1,1)$ & 2 & -2  & - & - & -4 & 0 & 1\\
			$c$ & 4 & $(-1,2)\times (-1,1)\times (-1,1)$ & 2 & 6  & - & - & - & - & 1\\
			\hline
			4 & 4 & $(0, -1)\times (0, 1)\times (2, 0)$& \multicolumn{7}{c|}{$x_A = \frac{4}{3}x_B = 8x_C = \frac{8}{3}x_D$}\\
			& & & \multicolumn{7}{c|}{$\beta^g_4=0$}\\
			& & & \multicolumn{7}{c|}{$\chi_1=4$, $\chi_2=\frac{2}{3}$, $\chi_3=4$}\\
			\hline
		\end{tabular}
	\end{center}
\end{table}
\begin{table}\scriptsize
	\caption{D6-brane configurations and intersection numbers of Model \ref{tb:model30}, and its MSSM gauge coupling relation is $g^2_a=\frac{5}{14}g^2_b=\frac{11}{6}g^2_c=\frac{11}{8}(\frac{5}{3}g^2_Y)=\frac{8}{63} 5^{3/4} \sqrt{11} \pi  e^{\phi ^4}$.}
	\label{tb:model30}
	\begin{center}
		\begin{tabular}{|c||c|c||c|c|c|c|c|c|c|c|}
			\hline\rm{model} \ref{tb:model30} & \multicolumn{10}{c|}{$U(4)\times U(2)_L\times U(2)_R\times USp(2)^2 $}\\
			\hline \hline			\rm{stack} & $N$ & $(n^1,l^1)\times(n^2,l^2)\times(n^3,l^3)$ & $n_{\Ysymm}$& $n_{\Yasymm}$ & $b$ & $b'$ & $c$ & $c'$ & 1 & 4\\
			\hline
			$a$ & 8 & $(-1,-1)\times (1,1)\times (1,1)$ & 0 & -4  & 6 & -3 & 0 & -3 & 1 & -1\\
			$b$ & 4 & $(-1,2)\times (-1,0)\times (5,1)$ & 9 & -9  & - & - & -10 & -9 & 0 & 1\\
			$c$ & 4 & $(-2,-1)\times (0,1)\times (1,1)$ & 1 & -1  & - & - & - & - & 1 & 0\\
			\hline
			1 & 2 & $(1, 0)\times (1, 0)\times (2, 0)$& \multicolumn{8}{c|}{$x_A = 22x_B = 2x_C = \frac{11}{5}x_D$}\\
			4 & 2 & $(0, -1)\times (0, 1)\times (2, 0)$& \multicolumn{8}{c|}{$\beta^g_1=-3$, $\beta^g_4=-3$}\\
			& & & \multicolumn{8}{c|}{$\chi_1=\frac{1}{\sqrt{5}}$, $\chi_2=\frac{11}{\sqrt{5}}$, $\chi_3=4 \sqrt{5}$}\\
			\hline
		\end{tabular}
	\end{center}
\end{table}
\begin{table}\scriptsize
	\caption{D6-brane configurations and intersection numbers of Model \ref{tb:model31}, and its MSSM gauge coupling relation is $g^2_a=\frac{1}{6}g^2_b=\frac{7}{6}g^2_c=\frac{35}{32}(\frac{5}{3}g^2_Y)=\frac{16}{135} 26^{3/4} \pi  e^{\phi ^4}$.}
	\label{tb:model31}
	\begin{center}
		\begin{tabular}{|c||c|c||c|c|c|c|c|c|c|c|}
			\hline\rm{model} \ref{tb:model31} & \multicolumn{10}{c|}{$U(4)\times U(2)_L\times U(2)_R\times USp(2)\times USp(4) $}\\
			\hline \hline			\rm{stack} & $N$ & $(n^1,l^1)\times(n^2,l^2)\times(n^3,l^3)$ & $n_{\Ysymm}$& $n_{\Yasymm}$ & $b$ & $b'$ & $c$ & $c'$ & 1 & 4\\
			\hline
			$a$ & 8 & $(-1,-1)\times (2,1)\times (1,1)$ & 0 & -8  & 0 & 3 & 0 & -3 & 1 & -2\\
			$b$ & 4 & $(4,-1)\times (1,0)\times (1,1)$ & -3 & 3  & - & - & 0 & 2 & 0 & 4\\
			$c$ & 4 & $(-2,-1)\times (-1,1)\times (1,1)$ & 2 & 6  & - & - & - & - & 1 & 2\\
			\hline
			1 & 4 & $(1, 0)\times (1, 0)\times (2, 0)$& \multicolumn{8}{c|}{$x_A = \frac{13}{2}x_B = \frac{13}{8}x_C = 26x_D$}\\
			4 & 2 & $(0, -1)\times (0, 1)\times (2, 0)$& \multicolumn{8}{c|}{$\beta^g_1=-3$, $\beta^g_4=4$}\\
			& & & \multicolumn{8}{c|}{$\chi_1=\sqrt{\frac{13}{2}}$, $\chi_2=2 \sqrt{26}$, $\chi_3=\frac{\sqrt{\frac{13}{2}}}{2}$}\\
			\hline
		\end{tabular}
	\end{center}
\end{table}
\begin{table}\scriptsize
	\caption{D6-brane configurations and intersection numbers of Model \ref{tb:model32}, and its MSSM gauge coupling relation is $g^2_a=\frac{26}{5}g^2_b=6g^2_c=2(\frac{5}{3}g^2_Y)=\frac{16}{5} \sqrt{6} \pi  e^{\phi ^4}$.}
	\label{tb:model32}
	\begin{center}
		\begin{tabular}{|c||c|c||c|c|c|c|c|c|c|}
			\hline\rm{model} \ref{tb:model32} & \multicolumn{9}{c|}{$U(4)\times U(2)_L\times U(2)_R\times USp(4) $}\\
			\hline \hline			\rm{stack} & $N$ & $(n^1,l^1)\times(n^2,l^2)\times(n^3,l^3)$ & $n_{\Ysymm}$& $n_{\Yasymm}$ & $b$ & $b'$ & $c$ & $c'$ & 3\\
			\hline
			$a$ & 8 & $(1,1)\times (-1,0)\times (-1,1)$ & 0 & 0  & 2 & 1 & 0 & -3 & 0\\
			$b$ & 4 & $(1,0)\times (2,-1)\times (1,3)$ & -5 & 5  & - & - & 16 & 8 & -1\\
			$c$ & 4 & $(-1,4)\times (0,1)\times (-1,1)$ & 3 & -3  & - & - & - & - & 1\\
			\hline
			3 & 4 & $(0, -1)\times (1, 0)\times (0, 2)$& \multicolumn{7}{c|}{$x_A = \frac{3}{2}x_B = \frac{1}{4}x_C = \frac{3}{2}x_D$}\\
			& & & \multicolumn{7}{c|}{$\beta^g_3=-4$}\\
			& & & \multicolumn{7}{c|}{$\chi_1=\frac{1}{2}$, $\chi_2=3$, $\chi_3=1$}\\
			\hline
		\end{tabular}
	\end{center}
\end{table}
\begin{table}\scriptsize
	\caption{D6-brane configurations and intersection numbers of Model \ref{tb:model33}, and its MSSM gauge coupling relation is $g^2_a=\frac{1}{6}g^2_b=\frac{11}{6}g^2_c=\frac{11}{8}(\frac{5}{3}g^2_Y)=\frac{8}{405} \sqrt{2} 161^{3/4} \pi  e^{\phi ^4}$.}
	\label{tb:model33}
	\begin{center}
		\begin{tabular}{|c||c|c||c|c|c|c|c|c|c|c|}
			\hline\rm{model} \ref{tb:model33} & \multicolumn{10}{c|}{$U(4)\times U(2)_L\times U(2)_R\times USp(2)^2 $}\\
			\hline \hline			\rm{stack} & $N$ & $(n^1,l^1)\times(n^2,l^2)\times(n^3,l^3)$ & $n_{\Ysymm}$& $n_{\Yasymm}$ & $b$ & $b'$ & $c$ & $c'$ & 1 & 4\\
			\hline
			$a$ & 8 & $(1,-1)\times (-1,1)\times (1,-1)$ & 0 & 4  & 3 & 0 & 0 & -3 & -1 & 1\\
			$b$ & 4 & $(1,-4)\times (1,0)\times (1,1)$ & 3 & -3  & - & - & -7 & 0 & 0 & 1\\
			$c$ & 4 & $(-2,1)\times (2,1)\times (-1,1)$ & -3 & -13  & - & - & - & - & -1 & -4\\
			\hline
			1 & 2 & $(1, 0)\times (1, 0)\times (2, 0)$& \multicolumn{8}{c|}{$x_A = 28x_B = \frac{28}{23}x_C = 7x_D$}\\
			4 & 2 & $(0, -1)\times (0, 1)\times (2, 0)$& \multicolumn{8}{c|}{$\beta^g_1=-3$, $\beta^g_4=1$}\\
			& & & \multicolumn{8}{c|}{$\chi_1=\sqrt{\frac{7}{23}}$, $\chi_2=\sqrt{161}$, $\chi_3=8 \sqrt{\frac{7}{23}}$}\\
			\hline
		\end{tabular}
	\end{center}
\end{table}

\FloatBarrier
\section{Gauge Coupling Unified Supersymmetric Pati-Salam Models}
\label{appB}

In this section, we present in particular some examples of supersymmetric Pati-Salam models with gauge unification at string scale.

\begin{table}[h]\scriptsize
	\caption{D6-brane configurations and intersection numbers of Model \ref{model1}, and its MSSM gauge coupling relation is $g^2_a=g^2_b=g^2_c=(\frac{5}{3}g^2_Y)=4 \sqrt{\frac{2}{3}} \pi  e^{\phi ^4}$.}
	\label{model1}
	\begin{center}
		\begin{tabular}{|c||c|c||c|c|c|c|c|c|c|c|c|c|}
			\hline\rm{model} \ref{model1} & \multicolumn{12}{c|}{$U(4)\times U(2)_L\times U(2)_R\times USp(2)^4 $}\\
			\hline \hline			\rm{stack} & $N$ & $(n^1,l^1)\times(n^2,l^2)\times(n^3,l^3)$ & $n_{\Ysymm}$& $n_{\Yasymm}$ & $b$ & $b'$ & $c$ & $c'$ & 1 & 2 & 3 & 4\\
			\hline
			$a$ & 8 & $(1,0)\times (-1,1)\times (-1,-1)$ & 0 & 0  & 3 & 0 & 0 & -3 & 0 & 0 & -1 & 1\\
			$b$ & 4 & $(1,-3)\times (-1,-1)\times (-1,0)$ & 2 & -2  & - & - & 0 & 0 & 0 & -3 & 1 & 0\\
			$c$ & 4 & $(1,-3)\times (-1,1)\times (0,-1)$ & 2 & -2  & - & - & - & - & -3 & 0 & 0 & 1\\
			\hline
			1 & 2 & $(1, 0)\times (2, 0)\times (1, 0)$& \multicolumn{10}{c|}{$x_A = x_B = \frac{1}{3}x_C = \frac{1}{3}x_D$}\\
			2 & 2 & $(1, 0)\times (0, -2)\times (0, 1)$& \multicolumn{10}{c|}{$\beta^g_1=-3$, $\beta^g_2=-3$, $\beta^g_3=-3$, $\beta^g_4=-3$}\\
			3 & 2 & $(0, -1)\times (2, 0)\times (0, 1)$& \multicolumn{10}{c|}{$\chi_1=\frac{1}{3}$, $\chi_2=1$, $\chi_3=2$}\\
			4 & 2 & $(0, -1)\times (0, 2)\times (1, 0)$& \multicolumn{10}{c|}{}\\
			\hline
		\end{tabular}
	\end{center}
\end{table}
\begin{table}[h]\scriptsize
	\caption{D6-brane configurations and intersection numbers of Model \ref{model2}, and its MSSM gauge coupling relation is $g^2_a=g^2_b=g^2_c=(\frac{5}{3}g^2_Y)=4 \sqrt{\frac{2}{3}} \pi  e^{\phi ^4}$.}
	\label{model2}
	\begin{center}
		\begin{tabular}{|c||c|c||c|c|c|c|c|c|c|c|c|c|}
			\hline\rm{model} \ref{model2} & \multicolumn{12}{c|}{$U(4)\times U(2)_L\times U(2)_R\times USp(2)^4 $}\\
			\hline \hline			\rm{stack} & $N$ & $(n^1,l^1)\times(n^2,l^2)\times(n^3,l^3)$ & $n_{\Ysymm}$& $n_{\Yasymm}$ & $b$ & $b'$ & $c$ & $c'$ & 1 & 2 & 3 & 4\\
			\hline
			$a$ & 8 & $(-1,1)\times (0,1)\times (-1,1)$ & 0 & 0  & 3 & 0 & 0 & -3 & -1 & 0 & 1 & 0\\
			$b$ & 4 & $(0,-1)\times (-3,-1)\times (-1,-1)$ & 2 & -2  & - & - & 0 & 0 & 1 & -3 & 0 & 0\\
			$c$ & 4 & $(1,0)\times (-3,-1)\times (-1,1)$ & 2 & -2  & - & - & - & - & 0 & 0 & 1 & -3\\
			\hline
			1 & 2 & $(1, 0)\times (1, 0)\times (2, 0)$& \multicolumn{10}{c|}{$x_A = 3x_B = x_C = 3x_D$}\\
			2 & 2 & $(1, 0)\times (0, -1)\times (0, 2)$& \multicolumn{10}{c|}{$\beta^g_1=-3$, $\beta^g_2=-3$, $\beta^g_3=-3$, $\beta^g_4=-3$}\\
			3 & 2 & $(0, -1)\times (1, 0)\times (0, 2)$& \multicolumn{10}{c|}{$\chi_1=1$, $\chi_2=3$, $\chi_3=2$}\\
			4 & 2 & $(0, -1)\times (0, 1)\times (2, 0)$& \multicolumn{10}{c|}{}\\
			\hline
		\end{tabular}
	\end{center}
\end{table}
\begin{table}[h]\scriptsize
	\caption{D6-brane configurations and intersection numbers of Model \ref{model3}, and its MSSM gauge coupling relation is $g^2_a=g^2_b=g^2_c=(\frac{5}{3}g^2_Y)=4 \sqrt{\frac{2}{3}} \pi  e^{\phi ^4}$.}
	\label{model3}
	\begin{center}
		\begin{tabular}{|c||c|c||c|c|c|c|c|c|c|c|c|c|}
			\hline\rm{model} \ref{model3} & \multicolumn{12}{c|}{$U(4)\times U(2)_L\times U(2)_R\times USp(2)^4 $}\\
			\hline \hline			\rm{stack} & $N$ & $(n^1,l^1)\times(n^2,l^2)\times(n^3,l^3)$ & $n_{\Ysymm}$& $n_{\Yasymm}$ & $b$ & $b'$ & $c$ & $c'$ & 1 & 2 & 3 & 4\\
			\hline
			$a$ & 8 & $(-1,1)\times (1,1)\times (-1,0)$ & 0 & 0  & 3 & 0 & 0 & -3 & 0 & -1 & 1 & 0\\
			$b$ & 4 & $(-1,-1)\times (-1,0)\times (1,-3)$ & 2 & -2  & - & - & 0 & 0 & 0 & 1 & 0 & -3\\
			$c$ & 4 & $(-1,1)\times (0,-1)\times (1,-3)$ & 2 & -2  & - & - & - & - & -3 & 0 & 1 & 0\\
			\hline
			1 & 2 & $(2, 0)\times (1, 0)\times (1, 0)$& \multicolumn{10}{c|}{$x_A = \frac{1}{3}x_B = \frac{1}{3}x_C = x_D$}\\
			2 & 2 & $(2, 0)\times (0, -1)\times (0, 1)$& \multicolumn{10}{c|}{$\beta^g_1=-3$, $\beta^g_2=-3$, $\beta^g_3=-3$, $\beta^g_4=-3$}\\
			3 & 2 & $(0, -2)\times (1, 0)\times (0, 1)$& \multicolumn{10}{c|}{$\chi_1=1$, $\chi_2=1$, $\chi_3=\frac{2}{3}$}\\
			4 & 2 & $(0, -2)\times (0, 1)\times (1, 0)$& \multicolumn{10}{c|}{}\\
			\hline
		\end{tabular}
	\end{center}
\end{table}
\begin{table}[h]\scriptsize
	\caption{D6-brane configurations and intersection numbers of Model \ref{model4}, and its MSSM gauge coupling relation is $g^2_a=g^2_b=g^2_c=(\frac{5}{3}g^2_Y)=4 \sqrt{\frac{2}{3}} \pi  e^{\phi ^4}$.}
	\label{model4}
	\begin{center}
		\begin{tabular}{|c||c|c||c|c|c|c|c|c|c|c|c|c|}
			\hline\rm{model} \ref{model4} & \multicolumn{12}{c|}{$U(4)\times U(2)_L\times U(2)_R\times USp(2)^4 $}\\
			\hline \hline			\rm{stack} & $N$ & $(n^1,l^1)\times(n^2,l^2)\times(n^3,l^3)$ & $n_{\Ysymm}$& $n_{\Yasymm}$ & $b$ & $b'$ & $c$ & $c'$ & 1 & 2 & 3 & 4\\
			\hline
			$a$ & 8 & $(1,1)\times (-1,1)\times (-1,0)$ & 0 & 0  & 0 & 3 & 0 & -3 & 0 & 1 & -1 & 0\\
			$b$ & 4 & $(-1,-1)\times (0,-1)\times (-1,-3)$ & -2 & 2  & - & - & 0 & 0 & 3 & 0 & -1 & 0\\
			$c$ & 4 & $(-1,-1)\times (-1,0)\times (1,-3)$ & 2 & -2  & - & - & - & - & 0 & 1 & 0 & -3\\
			\hline
			1 & 2 & $(2, 0)\times (1, 0)\times (1, 0)$& \multicolumn{10}{c|}{$x_A = \frac{1}{3}x_B = \frac{1}{3}x_C = x_D$}\\
			2 & 2 & $(2, 0)\times (0, -1)\times (0, 1)$& \multicolumn{10}{c|}{$\beta^g_1=-3$, $\beta^g_2=-3$, $\beta^g_3=-3$, $\beta^g_4=-3$}\\
			3 & 2 & $(0, -2)\times (1, 0)\times (0, 1)$& \multicolumn{10}{c|}{$\chi_1=1$, $\chi_2=1$, $\chi_3=\frac{2}{3}$}\\
			4 & 2 & $(0, -2)\times (0, 1)\times (1, 0)$& \multicolumn{10}{c|}{}\\
			\hline
		\end{tabular}
	\end{center}
\end{table}
\begin{table}[h]\scriptsize
	\caption{D6-brane configurations and intersection numbers of Model \ref{model5}, and its MSSM gauge coupling relation is $g^2_a=g^2_b=g^2_c=(\frac{5}{3}g^2_Y)=4 \sqrt{\frac{2}{3}} \pi  e^{\phi ^4}$.}
	\label{model5}
	\begin{center}
		\begin{tabular}{|c||c|c||c|c|c|c|c|c|c|c|c|c|}
			\hline\rm{model} \ref{model5} & \multicolumn{12}{c|}{$U(4)\times U(2)_L\times U(2)_R\times USp(2)^4 $}\\
			\hline \hline			\rm{stack} & $N$ & $(n^1,l^1)\times(n^2,l^2)\times(n^3,l^3)$ & $n_{\Ysymm}$& $n_{\Yasymm}$ & $b$ & $b'$ & $c$ & $c'$ & 1 & 2 & 3 & 4\\
			\hline
			$a$ & 8 & $(1,-1)\times (-1,0)\times (-1,-1)$ & 0 & 0  & 0 & 3 & -3 & 0 & 0 & -1 & 0 & 1\\
			$b$ & 4 & $(0,-1)\times (-1,-3)\times (-1,-1)$ & -2 & 2  & - & - & 0 & 0 & 3 & -1 & 0 & 0\\
			$c$ & 4 & $(-1,0)\times (1,3)\times (-1,1)$ & -2 & 2  & - & - & - & - & 0 & 0 & 3 & -1\\
			\hline
			1 & 2 & $(1, 0)\times (1, 0)\times (2, 0)$& \multicolumn{10}{c|}{$x_A = \frac{1}{3}x_B = x_C = \frac{1}{3}x_D$}\\
			2 & 2 & $(1, 0)\times (0, -1)\times (0, 2)$& \multicolumn{10}{c|}{$\beta^g_1=-3$, $\beta^g_2=-3$, $\beta^g_3=-3$, $\beta^g_4=-3$}\\
			3 & 2 & $(0, -1)\times (1, 0)\times (0, 2)$& \multicolumn{10}{c|}{$\chi_1=1$, $\chi_2=\frac{1}{3}$, $\chi_3=2$}\\
			4 & 2 & $(0, -1)\times (0, 1)\times (2, 0)$& \multicolumn{10}{c|}{}\\
			\hline
		\end{tabular}
	\end{center}
\end{table}
\begin{table}[h]\scriptsize
	\caption{D6-brane configurations and intersection numbers of Model \ref{model6}, and its MSSM gauge coupling relation is $g^2_a=g^2_b=g^2_c=(\frac{5}{3}g^2_Y)=4 \sqrt{\frac{2}{3}} \pi  e^{\phi ^4}$.}
	\label{model6}
	\begin{center}
		\begin{tabular}{|c||c|c||c|c|c|c|c|c|c|c|c|c|}
			\hline\rm{model} \ref{model6} & \multicolumn{12}{c|}{$U(4)\times U(2)_L\times U(2)_R\times USp(2)^4 $}\\
			\hline \hline			\rm{stack} & $N$ & $(n^1,l^1)\times(n^2,l^2)\times(n^3,l^3)$ & $n_{\Ysymm}$& $n_{\Yasymm}$ & $b$ & $b'$ & $c$ & $c'$ & 1 & 2 & 3 & 4\\
			\hline
			$a$ & 8 & $(-1,1)\times (-1,-1)\times (1,0)$ & 0 & 0  & 3 & 0 & -3 & 0 & 0 & -1 & 1 & 0\\
			$b$ & 4 & $(-1,-1)\times (-1,0)\times (1,-3)$ & 2 & -2  & - & - & 0 & 0 & 0 & 1 & 0 & -3\\
			$c$ & 4 & $(1,1)\times (0,1)\times (-1,-3)$ & -2 & 2  & - & - & - & - & 3 & 0 & -1 & 0\\
			\hline
			1 & 2 & $(2, 0)\times (1, 0)\times (1, 0)$& \multicolumn{10}{c|}{$x_A = \frac{1}{3}x_B = \frac{1}{3}x_C = x_D$}\\
			2 & 2 & $(2, 0)\times (0, -1)\times (0, 1)$& \multicolumn{10}{c|}{$\beta^g_1=-3$, $\beta^g_2=-3$, $\beta^g_3=-3$, $\beta^g_4=-3$}\\
			3 & 2 & $(0, -2)\times (1, 0)\times (0, 1)$& \multicolumn{10}{c|}{$\chi_1=1$, $\chi_2=1$, $\chi_3=\frac{2}{3}$}\\
			4 & 2 & $(0, -2)\times (0, 1)\times (1, 0)$& \multicolumn{10}{c|}{}\\
			\hline
		\end{tabular}
	\end{center}
\end{table}
\begin{table}[h]\scriptsize
	\caption{D6-brane configurations and intersection numbers of Model \ref{model7}, and its MSSM gauge coupling relation is $g^2_a=g^2_b=g^2_c=(\frac{5}{3}g^2_Y)=4 \sqrt{\frac{2}{3}} \pi  e^{\phi ^4}$.}
	\label{model7}
	\begin{center}
		\begin{tabular}{|c||c|c||c|c|c|c|c|c|c|c|c|c|}
			\hline\rm{model} \ref{model7} & \multicolumn{12}{c|}{$U(4)\times U(2)_L\times U(2)_R\times USp(2)^4 $}\\
			\hline \hline			\rm{stack} & $N$ & $(n^1,l^1)\times(n^2,l^2)\times(n^3,l^3)$ & $n_{\Ysymm}$& $n_{\Yasymm}$ & $b$ & $b'$ & $c$ & $c'$ & 1 & 2 & 3 & 4\\
			\hline
			$a$ & 8 & $(1,0)\times (-1,1)\times (-1,-1)$ & 0 & 0  & 3 & 0 & -3 & 0 & 0 & 0 & -1 & 1\\
			$b$ & 4 & $(1,-3)\times (-1,-1)\times (-1,0)$ & 2 & -2  & - & - & 0 & 0 & 0 & -3 & 1 & 0\\
			$c$ & 4 & $(-1,-3)\times (1,1)\times (0,1)$ & -2 & 2  & - & - & - & - & 3 & 0 & 0 & -1\\
			\hline
			1 & 2 & $(1, 0)\times (2, 0)\times (1, 0)$& \multicolumn{10}{c|}{$x_A = x_B = \frac{1}{3}x_C = \frac{1}{3}x_D$}\\
			2 & 2 & $(1, 0)\times (0, -2)\times (0, 1)$& \multicolumn{10}{c|}{$\beta^g_1=-3$, $\beta^g_2=-3$, $\beta^g_3=-3$, $\beta^g_4=-3$}\\
			3 & 2 & $(0, -1)\times (2, 0)\times (0, 1)$& \multicolumn{10}{c|}{$\chi_1=\frac{1}{3}$, $\chi_2=1$, $\chi_3=2$}\\
			4 & 2 & $(0, -1)\times (0, 2)\times (1, 0)$& \multicolumn{10}{c|}{}\\
			\hline
		\end{tabular}
	\end{center}
\end{table}

\end{document}